\documentclass[preprint,english,letterpaper,showpacs,floatfix]{revtex4}
\usepackage[latin1]{inputenc}
\usepackage{amsmath,amssymb}
\usepackage{epsfig}
\usepackage{pgf}

\makeatletter

\begin{document}
\title{Shedding Some New Lights upon the Stellar Quasi-Normal Modes}

\author{E. Abdalla}
\email{eabdalla@fma.if.usp.br}

\author{D. Giugno}
\email{dgiugno@fma.if.usp.br}

\affiliation{Instituto de F\'{\i}sica, Universidade de S\~{a}o Paulo \\
C.P. 66318, CEP 05315, S\~{a}o Paulo-SP, Brazil}



\begin{abstract}

In the current paper we present some new data on the issue of
quasi-normal modes (QNMs) of uniform, neutron and quark stars. These 
questions have already been addressed in the literature before, but we 
have found some interesting features that have not been discussed so
far. We have increased the range of frequency values for the scalar
and axial perturbations of such stars and made a comparison between
such QNMs and those of the very well-known Schwarzschild black holes. 
Also addressed in this work was the interesting feature of competing 
modes, which appear not only for uniform stars, but for quark stars as well.

\end{abstract}

\maketitle

\section{Introduction}

Quasi normal modes (QNM's) have been studied for quite a long time due
to the possibility of gathering information from astrophysical objects
in terms of their response to external perturbations, in the same
sense as we may study a bell from its sound. For reviews and earlier
notes see \cite{Chand-Ferr}, \cite{Kokkotas}, \cite{Price}. They are
particularly useful to grasp also general properties of the metric
under consideration - see \cite{WMA}, \cite{WLA}, \cite{Kar}.

Stellar QNM's have been under considerable scrutiny over the last
decades, since they provide not only a test for Einstein's theory of
General Relativity (GR), but also a look into the stellar structure
and, indirectly, into the very nature of stellar matter and its
properties, such as its equation of state (EOS).

Stars have an internal structure which must be accounted for when
studying any kind of perturbation, be it a test scalar field or a 
gravitational perturbation. Among these, the axial perturbations are 
much easier to compute, since they do not mix with the fluid modes 
of the star \cite{Chand-FerrB}. Such axial perturbations (along 
with the scalar and electromagnetic) can be
described by a wave equation of the form
\begin{equation}
-\frac{\partial^2W}{\partial t^2}+\frac{\partial^2W}{\partial x^2}=VW,
\end{equation}
where $W$ stands for the amplitude and $V$, for the perturbative
potential. For details, see the Appendix.

The polar perturbations mix with the fluid modes and require a
completely different approach in their numerical analysis and
evolution, and are not dealt with here.

Uniform stars, that is, stars possessing a uniform density
$\varepsilon=\varepsilon_{0}$ are idealized astrophysical objects,
since they cannot exist in nature. Unrealistic as they may
sound, though, such stars provide an interesting background in
which some insight on the field dynamics in stellar geometries may be
gained, since the physical quantities of relevance (mass, pressure and
gravitational potentials of the metric) are very straightforward to
evaluate. We begin with such stars, but we do not limit ourselves to them. 
We present some results for the neutron and quark stars and compare these 
results to those of the well-known Schwarzschild black holes. 

For the neutron stars, we have dealt with the simplest model available, that 
of Oppenheimer and Volkov \cite{OV-39}, consisting of a pure Fermi gas of neutrons. 
For the quark stars, we have also used one of the simplest models available, the MIT Bag Model.

In this paper we consider QNMs of uniform, neutron and quark stars and compare 
with the analogous results for black holes, in an attempt to describe properties 
inherent of astrophysical objects.

In section II we present some results about uniform stars. Within that 
section, the question of secondary modes in such stars is discussed. 
In section III, we discuss neutron star QNMs and in section IV we do 
the same for quark stars. Comparative charts for all QNMs are available 
in section V and the remarks and conclusions are left for section VI. 

Concerning units, we have used the geometric system of units, for which 
$\hbar=c=G=1$. This means that the masses have dimension of length and 
are measured in metres. The conversion factor from metres to kilograms 
is $c^2/G$. Before proceeding, we just recall a definition which will 
be very useful throughout this paper, that of \emph{compactness} of a 
star. The compactness $c$ for a spherically symmetric star is defined 
in the literature as
\begin{equation}
c=\frac{r_{g}}{R},
\end{equation} 
in which $r_{g}=2M$ is the star's gravitational radius and $R$ is its actual radius. 
\section{QNMs of Uniform Stars - Some Results}

We begin with a series of figures - from the data we have tabulated - on the frequencies of the QNMs for
uniform stars, with some of the masses we have chosen for neutron
and quark stars, namely $M=1048m$, $M=977m$, $M=665m$ and $M=330m$ (the first 
two were also used for neutron and quark stars), for the sake of comparison. 
One must bear in mind that such values are not special in any way, being just 
the results of star integrations for some particular choices of the central 
density $\varepsilon_{0}$, and even these latter choices are just choices - 
for more details on the matter, see section III. In what follows, $c$ stands for
the compactness of the star, $\ell$ for the
multipole index and $\omega_{R}$ and $\omega_{I}$ for the real and the
imaginary part of the frequency, respectively. 
\begin{figure}[h]
\begin{center}
\rotatebox{-90}{\mbox{\epsfig{file=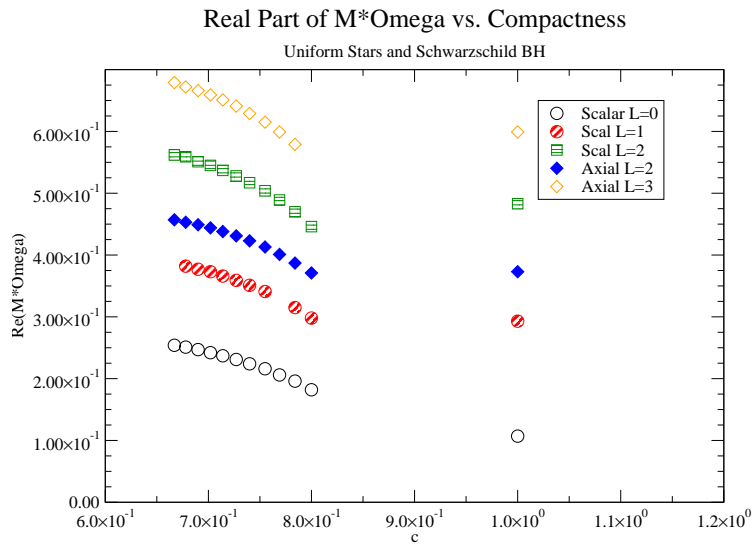,width=0.5\linewidth,clip=}}}
\end{center}
\caption{QNM frequencies (real part) for a uniform star, various fields, changing
compactness. The $c=1$ mark is for the Schwarzschild BH.}
\label{UniA}
\end{figure}

\begin{figure}[h]
\begin{center}
\rotatebox{-90}{\mbox{\epsfig{file=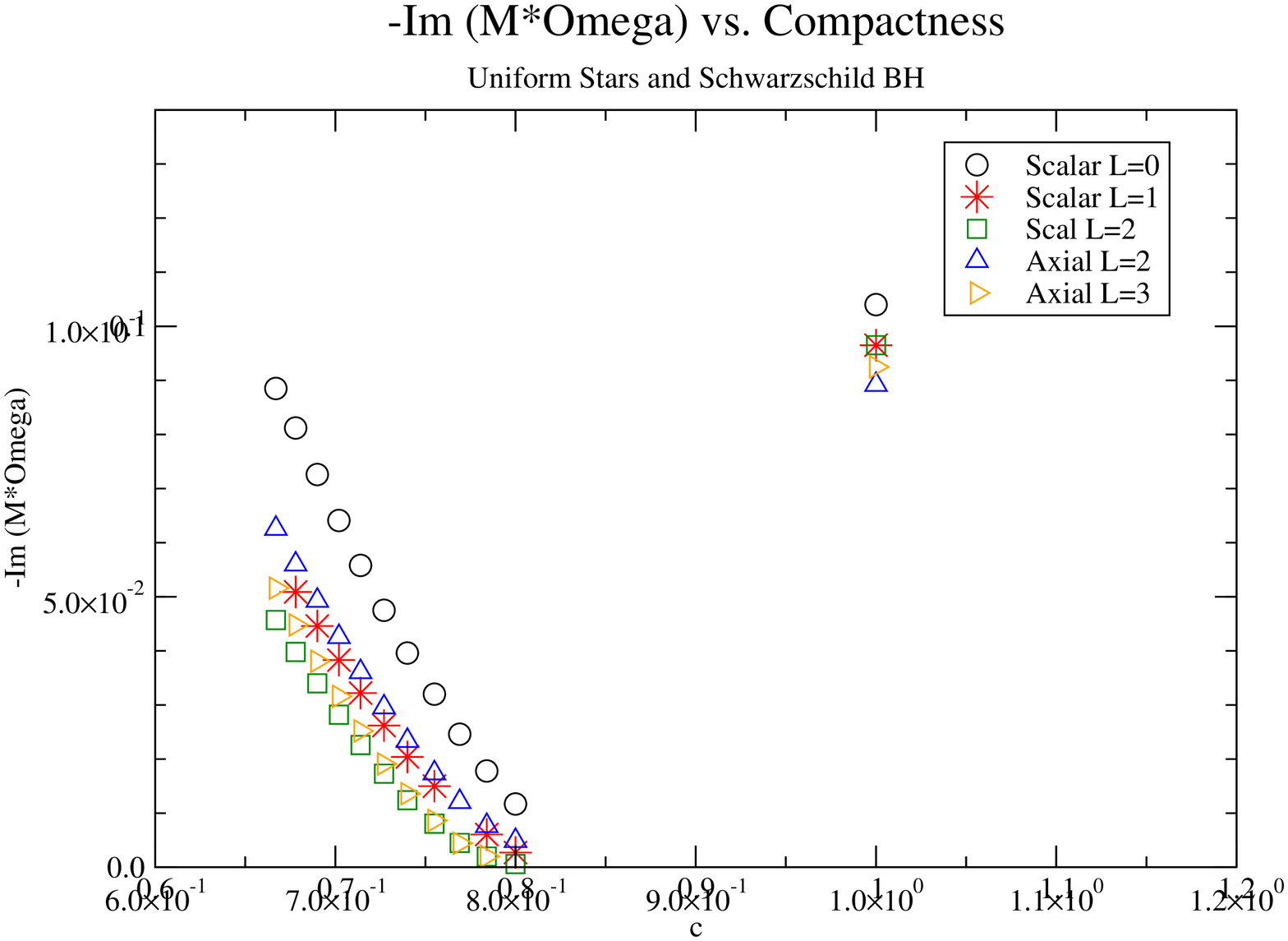,width=0.5\linewidth,clip=}}}
\end{center}
\caption{QNMs frequencies (negative of the imaginary part) for a
uniform star, various fields, changing compactness. The $c=1$ mark is
for the Schwarzschild BH.}
\label{UniB}
\end{figure}

From the data we have compiled, we can make a few initial remarks. 
First, we have noticed that $\omega\propto\frac{1}{M}$, as in the pure 
Schwarzschild case. Such a scaling property for $\omega$ in uniform 
stars was also explored in this work, and more data are left for subsection B. Second,
increases of $\omega_{R}$ with $\ell$ (as expected). Moreover, all axial
frequencies have smaller real parts than their scalar counterparts,
given some $\ell$ (exactly like Schwarzschild). But $-\omega_{I}$ is
higher for axial perturbations than for scalar ones (in contrast to the
Schwarzschild case). 

\begin{figure}[h]
\begin{center}
\rotatebox{-90}{\mbox{\epsfig{file=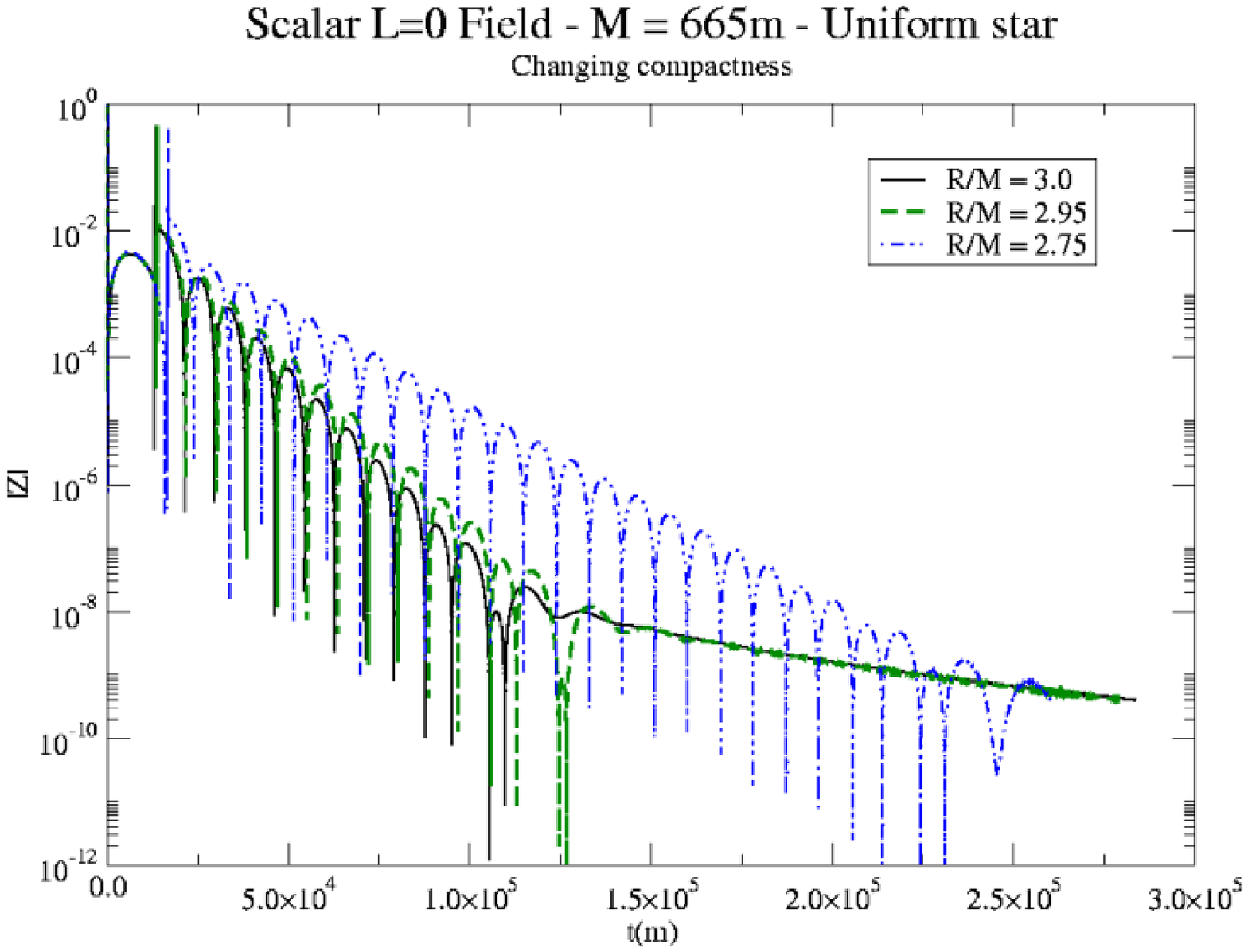,width=0.6\linewidth,clip=}}}  \end{center}
\caption{QNMs for a uniform compact star. Scalar $\ell=0$ case shown. $M=665m$.}
\label{L0b}
\end{figure}

We shall illustrate our data with a set of graphics. We begin with
the picture for the scalar $\ell=0$ case, which can be viewed in Fig. (\ref{L0b}). 

\begin{figure}[h]
\begin{center}
\rotatebox{-90}{\mbox{\epsfig{file=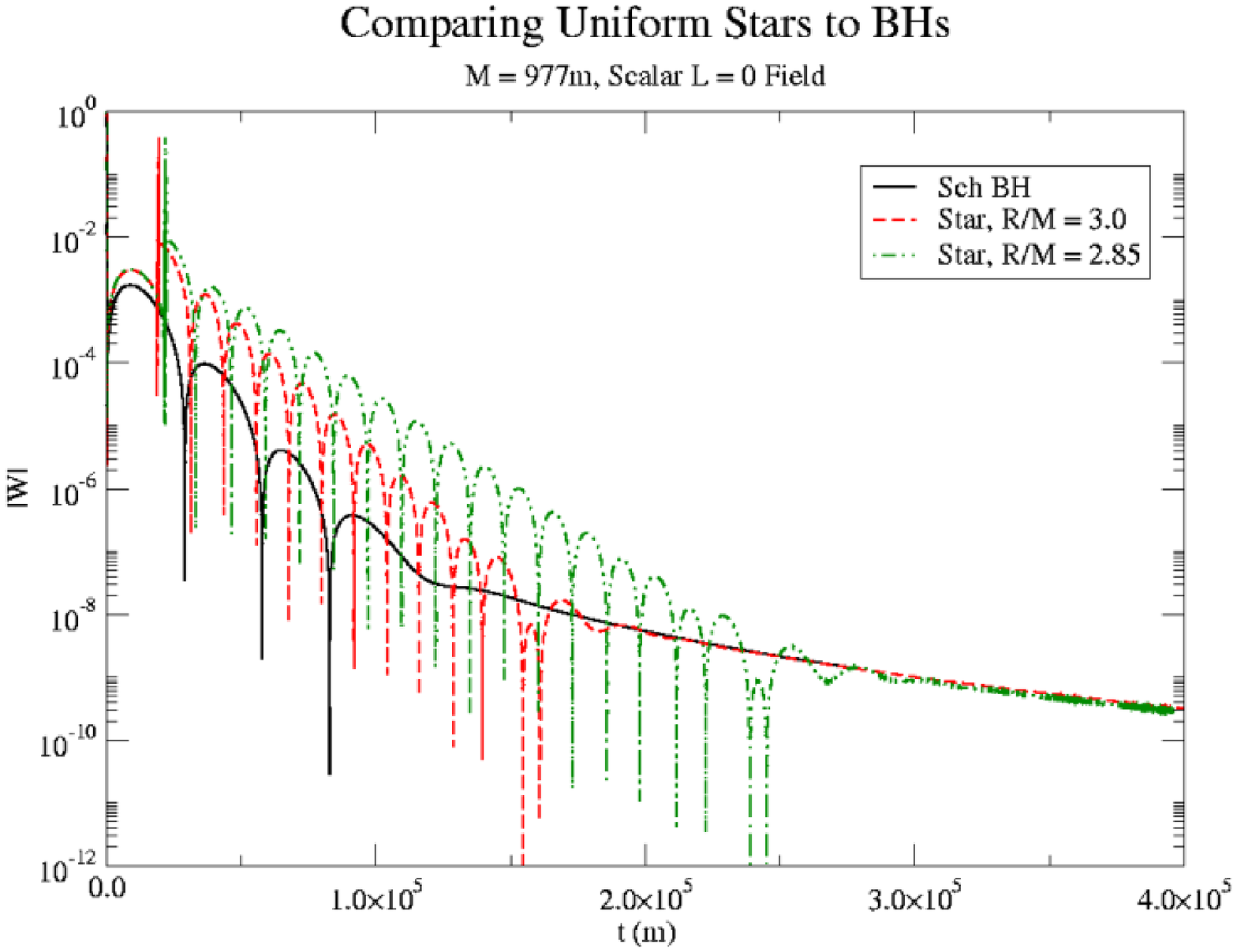,width=0.6\linewidth,clip=}}}\end{center}
\caption{QNMs and tails for stars and Schwarzschild BHs. Scalar
  $\ell=0$ case shown. $M=977m$. The tails have the same qualitative behaviour.}\label{CompBHUNI}
\end{figure}

In Fig. (\ref{L0b}), the tail obeys a power-law, namely $t^{-4}$. A
comparison between uniform star and Schwarzschild BH modes for the
$\ell=0$ scalar field is provided in
Fig. (\ref{CompBHUNI}). Notice that the BH-background scalar field
oscillates much less than in uniform stars, although the tail decays
much in the same way, according to the same power law.

\subsection{QNM Overtones}

We have detected, throughout our investigation of compact uniform
stars, the presence of secondary modes which decay faster than the
dominant ones. Since we have seen a similar behavior in the
Schwarzschild BH context (see note at the end of this subsection) and
we have checked these Schwarzschild BH data and concluded that they
correspond to the first overtones, we can speak of overtones of the
fundamental modes in the stellar context, also. 

\begin{figure}[h]
\begin{center}
\epsfig{file=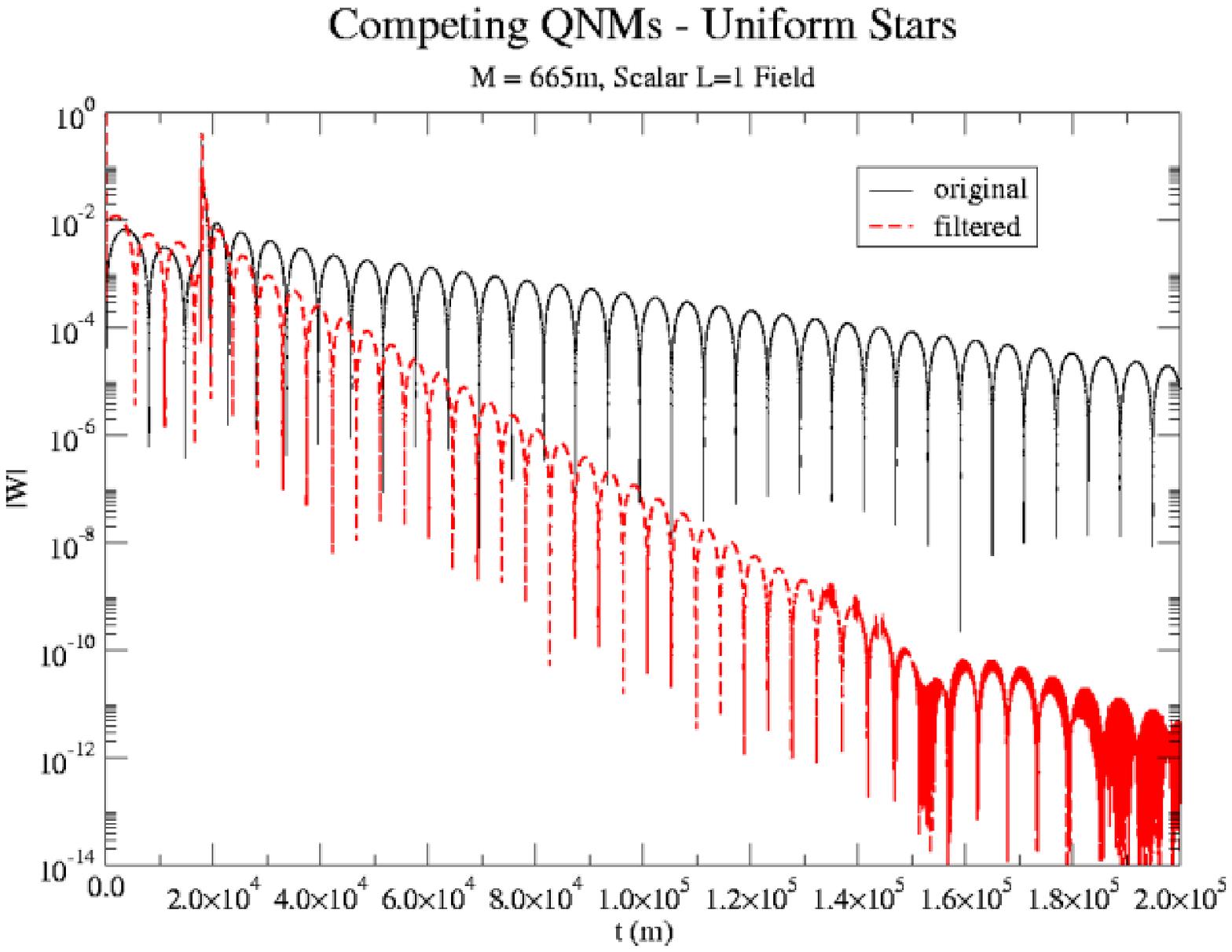,width=0.6\linewidth,clip=}
\end{center}
\caption{Competing QNMs for a uniform compact star, $c=0.741$. Scalar
$\ell=1$, $M=665m$ case shown. The secondary mode decays much faster.}
\label{competing1}
\end{figure}

\begin{figure}[h]
\begin{center}
\epsfig{file=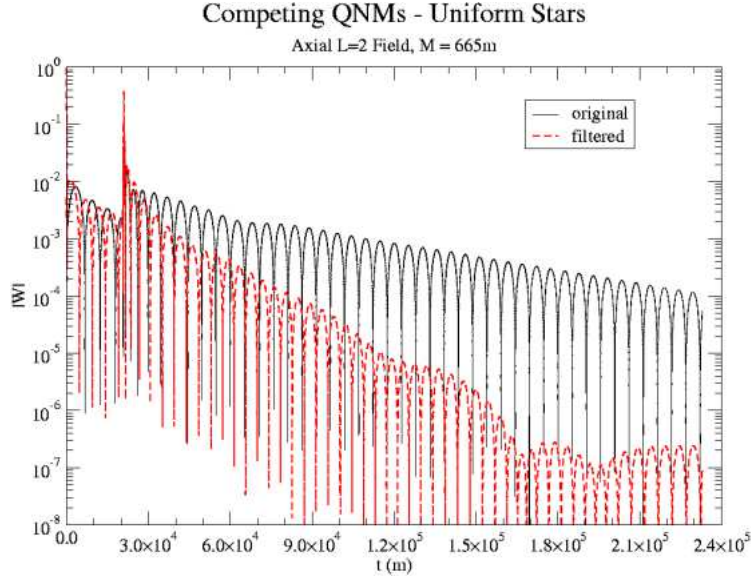,width=0.6\linewidth,clip=}
\end{center}
\caption{Competing QNMs for a uniform compact star, $c=0.769$. Axial
$\ell=2$, $M=665m$ case shown.}
\label{competing2}
\end{figure}

\begin{figure}[h]
\begin{center}
\epsfig{file=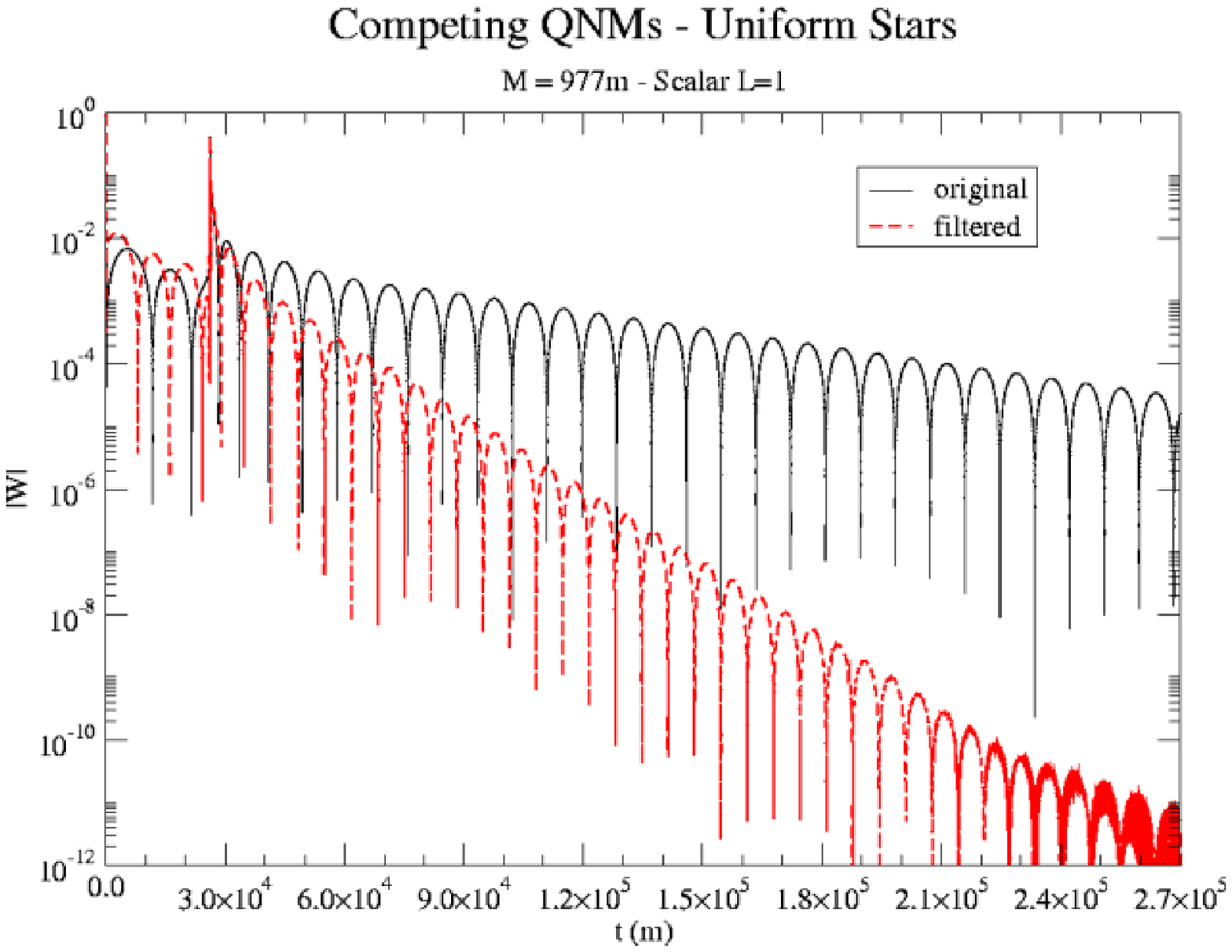,width=0.6\linewidth,clip=}
\end{center}
\caption{Competing QNMs for a uniform compact star, $c=0.741$. Scalar
$\ell=1$, $M=977m$ case shown.}
\label{competing3}
\end{figure}

These overtones show themselves in the form of wiggles in the
envelopes of the Misner curves characterising the QNMs, as if there
were modes (actually, the overtones) competing with the dominant ones. 
This issue of competing modes deserves special attention, especially for
- but not limiting to - high compactnesses ($c>0.76$), when these
wiggles become clearer. Three examples
of competing modes are provided in Figs. (\ref{competing1}),
(\ref{competing2}) and (\ref{competing3}).

A comment is due on the method of extraction of such secondary modes:
we have fitted a damped oscillating function to the original data, so
that we could know the frequency $\omega$ of the dominant mode (the
one with the weaker damping) - visible at the end of the time
evolution in all figures - and subtracted the fitted function from the
original data. The remaining mode is the competing mode we have just talked
about. For some extra details on the method, see \cite{dgiugno-06}.

Our fittings are not always very precise, so that we give no more than
3 significant figures for them, in the dominant mode. The remaining
mode is not always very easy to fit, primarily due to these precision
limitations. This remaining mode may look as if it were trembling, but
in some cases we can still made a (quite poor) fitting to it. We
expect it to have a stronger - usually much stronger - decay, which
was indeed the case. 

In the few cases in which we have seen a somewhat clear secondary
mode, we could notice also a slightly faster oscillation. For the
modes seen on Fig. (\ref{competing1}) we have, for instance,
$M\omega=0.351-0.0204i$ for the dominant mode
and $M\omega=0.47-0.087i$ for the secondary
mode, whereas for those seen on Fig. (\ref{competing2}) we have
$M\omega=0.401-0.0120i$  for the dominant mode
and $M\omega=0.49-0.042i$ for the competing
mode, so that the latter has a much stronger decay. The number of
significant figures for the latter was also reduced, since it was
obtained after a three-figure fitting had been performed. More data
are available in Table (\ref{TableCompeting}), for
$\frac{R}{M}=2.6$. 

These data simply confirm what we have said about
the secondary modes in question: they oscillate somewhat faster (in contrast 
to their Schwarzschild counterparts) and
decay much faster than the dominant ones (similarly to the
Schwarzschild BH case). This behaviour is confirmed in Table
(\ref{TableCompeting2}), where the relation between the real and
imaginary parts of $\omega$ for the dominant and the secondary modes
is shown (for the scalar $\ell=1$ field and $M=977m$). The
oscillations are about 30\% faster for the secondary modes and this
percentage seems to have a weak - if any - correlation to the
compactness $c$, but the damping rates can be at least 4 times
higher for the secondary modes, and sometimes up to 9 times,
increasing sharply with $c$. For the Schwarzschild BH, at least for
the $\ell=2$ and $\ell=3$ axial cases, the decay rate was around 3
times faster for the competing mode (the first overtone). As for the
first overtones, we could see that $\omega_{R}\propto\frac{1}{M}$, but
for $\omega_{I}$ the data seemed to fluctuate somewhat. This may be
due to the fitting precision limitations to which we have referred
earlier. See table (\ref{TableCompeting}).

\begin{table}[tbp]
\begin{tabular}{|c|c|c|c|c|}
\hline
$Mass(m)$ & $field$ & $\ell$ & $M\omega^{DOM}$ & $M\omega^{SEC}_{I}$ \\
\hline
$1048$ & $scal$ & $1$ & $0.329-0.0102i$ & $0.42-0.056i$ \\
\hline
$1048$ & $axial$ & $2$ & $0.401-0.0121i$ & $0.49-0.046i$ \\
\hline
$977$ & $scal$ & $1$ & $0.329-0.0102i$ & $0.42-0.053i$ \\
\hline
$977$ & $axial$ & $2$ & $0.402-0.0120i$ & $0.49-0.050i$ \\
\hline
$665$ & $scal$ & $1$ & $0.329-0.0102i$ & $0.43-0.055i$ \\
\hline
$665$ & $axial$ & $2$ & $0.401-0.0120i$ & $0.49-0.042i$ \\
\hline
\end{tabular}
\caption{Comparing dominant ($\omega^{DOM}$) and competing
($\omega^{SEC}$) mode frequencies for several masses of uniform stars
with $R/M=2.6$ ($c=0.769$).}
\label{TableCompeting}
\end{table}


\begin{table}[tbp]
\begin{tabular}{|c|c|c|c|c|c|c|}
\hline
$c$ & $M\omega^{DOM}_{R}$ & $-M\omega^{DOM}_{I}$ & $M\omega^{SEC}_{R}$ & 
$-M\omega^{DOM}_{I}$ & $\frac{\omega^{SEC}_{R}}{\omega^{DOM}_{R}}$ & 
$\frac{\omega^{SEC}_{I}}{\omega^{DOM}_{I}}$ \\
\hline
$0.727$ & $0.359$ & $0.0262$ & $0.48$ & $0.11$ & $1.3$ & $4.2$ \\
\hline
$0.741$ & $0.351$ & $0.0204$ & $0.47$ & $0.087$ & $1.3$ & $4.3$ \\
\hline
$0.755$ & $0.341$ & $0.0150$ & $0.45$ & $0.073$ & $1.3$ & $4.9$ \\
\hline
$0.769$ & $0.329$ & $0.0102$ & $0.43$ & $0.055$ & $1.3$ & $5.4$ \\
\hline
$0.784$ & $0.316$ & $0.00608$ & $0.40$ & $0.042$ & $1.3$ & $6.9$ \\
\hline
$0.800$ & $0.298$ & $0.00272$ & $0.39$ & $0.025$ & $1.3$ & $9.2$ \\
\hline
\end{tabular}
\caption{Comparing dominant and secondary modes for the $\ell=1$ scalar field, 
for a uniform star with mass $M=665m$.}
\label{TableCompeting2}
\end{table}

A brief remark on the Schwarzschild BH case: it is well-known
\cite{Kokkotas} that in their context, secondary modes (or first overtones) may indeed
appear. The same holds for higher overtones. We have searched for them in the same context in
order to test the procedure we have adopted to extract secondary modes
from uniform compact stars. The data we got for the axial $\ell=2$
axial field, for instance, were compared to those of \cite{Kokkotas}
and we have found a good agreement between our results and
theirs, indicating that our method, however simple-minded as it seems,
may indeed yield interesting results. To be more precise, for the
$n=0$ mode (the fundamental), they got $M\omega=0.37367-0.08896i$,
the same as ours. For the first overtone ($n=1$)
they had $M\omega=0.34671-0.27391i$ and we had
$M\omega=0.352-0.272i$ for $M=1048m$. This agreement is not so good as
that for the $n=0$ mode, but seems to be good enough for us,
indicating that our numerical procedures are on the right track. No
higher overtones ($n>1$) were detected in the present context. Again, see 
\cite{dgiugno-06} for closer details.

From the last paragraph, we may conclude that the first overtones decay 
faster in both the Schwarzschild case and the uniform compact star case, 
though in the latter case the decay rate depends on the compactness, being 
higher in more compact stars and usually higher than in the Schwarzschild context. 
One important difference between first overtones in Schwarzschild BHs and uniform 
stars backgrounds concerns their oscillations, which are slightly slower than 
in the fundamental mode for the former and somewhat faster (though not much) 
in the latter.

\subsection{Scaling Properties and Other Comments}

\begin{figure}[h]
\begin{center}
\rotatebox{-90}{\mbox{\epsfig{file=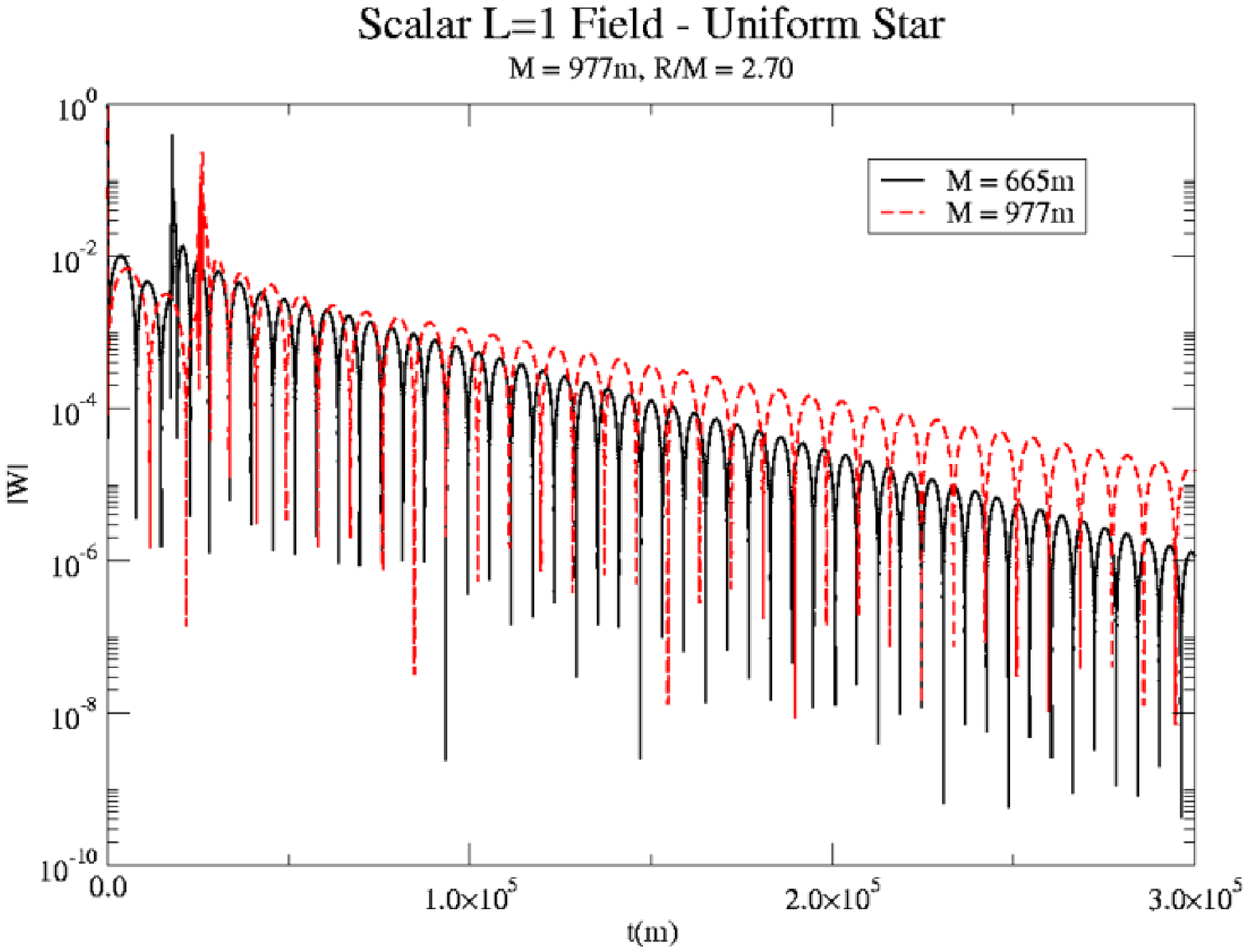,width=0.6\linewidth,clip=}}}
\end{center}
\caption{QNMs for a uniform compact star. The field oscillates and
decays more
slowly for a more massive star, exactly as one would expect from a
Schwarzschild black hole.}
\label{ChMass1}
\end{figure}

We can also check the dependence of the modes - for a given
compactness and perturbation - on the mass, as shown in
Fig. (\ref{ChMass1}). 

Besides fig. (\ref{ChMass1}), one may find an interesting scaling property 
for uniform star QNMs.  

\begin{table}[tbp]
\begin{tabular}{|c|c|c|c|c|}
\hline
$Mass(m)$ & $c$ & $field$ & $\ell$ & $M\omega$ \\
\hline
$1048$ & $0.769$ & $scal$ & $1$ & $0.329-0.0102i$ \\
\hline
$1048$ & $0.769$ & $axial$ & $2$ & $0.401-0.0121i$ \\
\hline
$977$ & $0.769$ & $scal$ & $1$ & $0.329-0.0102i$ \\
\hline
$977$ & $0.769$ & $axial$ & $2$ & $0.402-0.0120i$ \\
\hline
$665$ & $0.769$ & $scal$ & $1$ & $0.329-0.0102i$ \\
\hline
$665$ & $0.769$ & $axial$ & $2$ & $0.401-0.0120i$ \\
\hline
$1048$ & $0.740$ & $axial$ & $2$ & $0.422-0.0233i$ \\
\hline
$977$ & $0.740$ & $axial$ & $2$ & $0.421-0.0234i$ \\
\hline
$665$ & $0.740$ & $axial$ & $2$ & $0.423-0.0233i$ \\
\hline
$977$ & $0.769$ & $scalar$ & $2$ & $0.489-0.00440i$ \\
\hline
$665$ & $0.769$ & $scalar$ & $2$ & $0.489-0.00448i$ \\
\hline
\end{tabular}
\caption{Search for scaling properties of the QNM frequencies - Uniform Stars.}
\label{MassScaling1}
\end{table}

The table (\ref{MassScaling1}) shows that for a given field,
compactness and $\ell$, the quantity $M\omega$ is practically
constant. For instance, when $\ell=1$ for the scalar field and
$c=0.769$ ($R/M=2.6$), one has $M\omega$ being $0.329-0.0102i$ for
both $M=977m$ and $M=665m$, while for $c=0.740$ ($R/M=2.7$) one has
$M\omega=0.351-0.0204i$ for $M=665m$, and almost the same 
($M\omega=0.349-0.0205i$) for $M=977m$. This can be corroborated 
for a larger mass spectrum, different compactnesses, fields and $\ell$ values. 

\begin{figure}[h]
\begin{center}
\rotatebox{-90}{\mbox{\epsfig{file=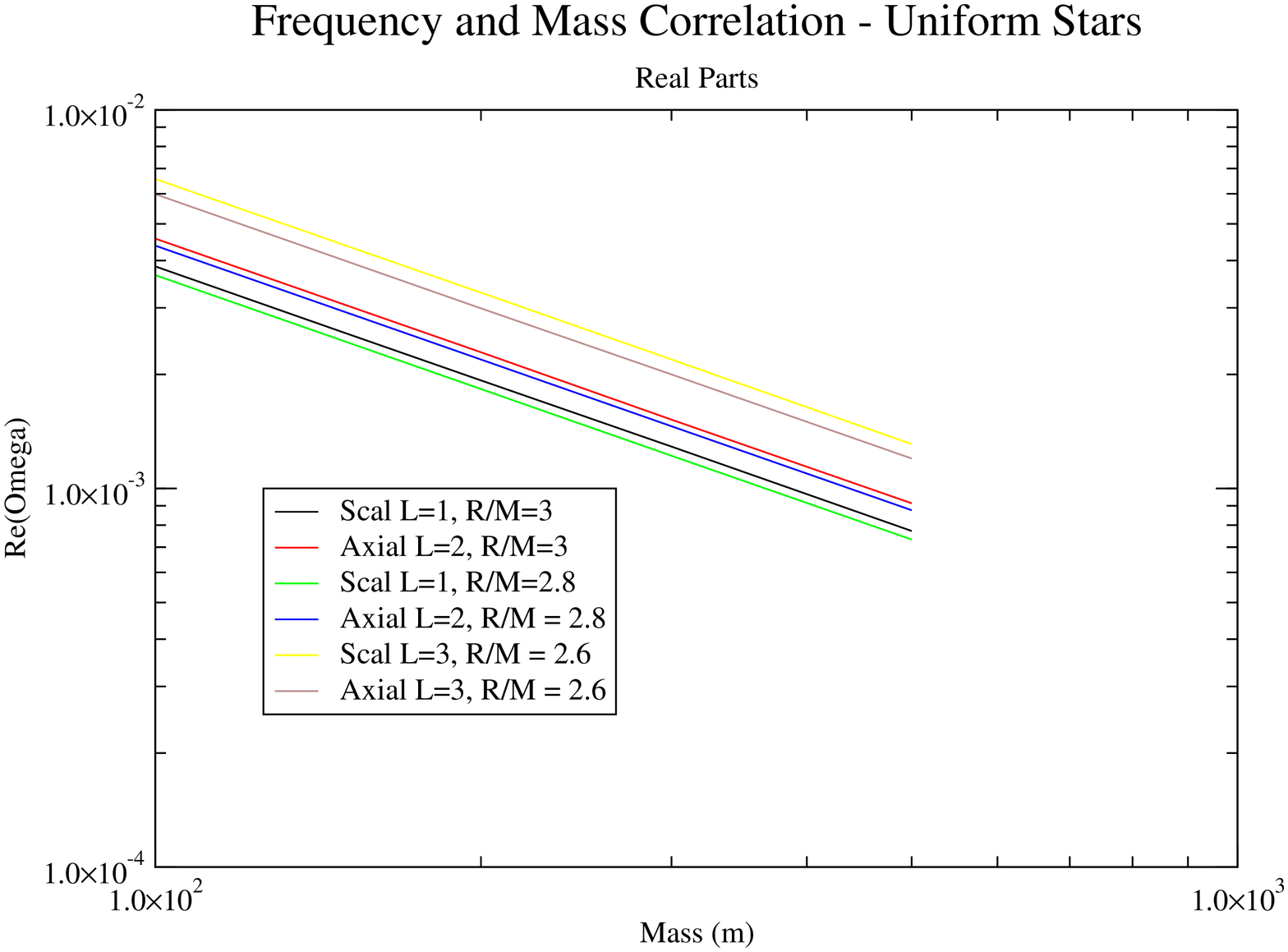,width=0.55\linewidth,clip=}}}
\end{center}
\caption{$\omega_{R}$ vs. the mass.}
\label{FreqTable1}
\end{figure}

\begin{figure}[h]
\begin{center}
\rotatebox{-90}{\mbox{\epsfig{file=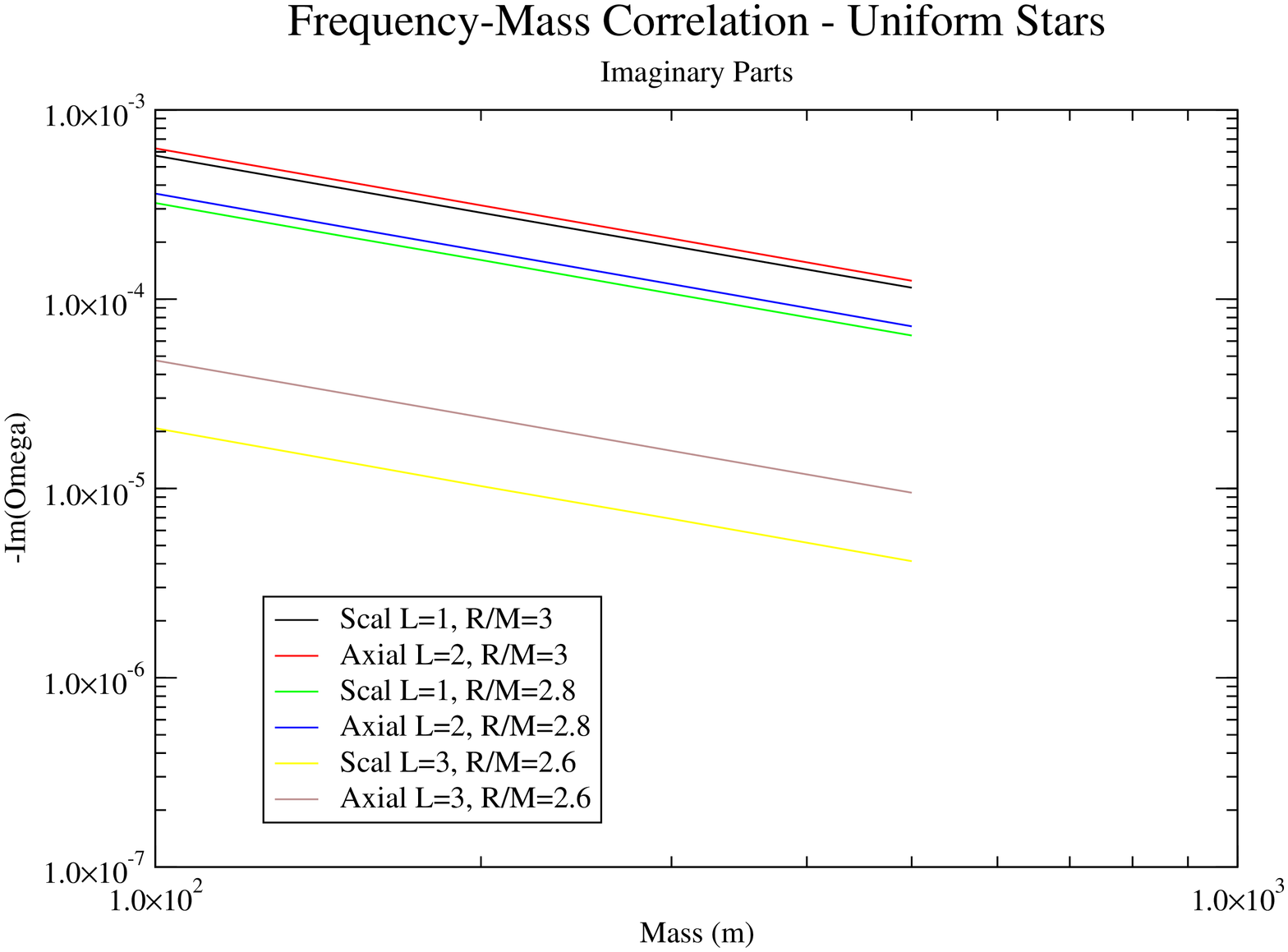,width=0.55\linewidth,clip=}}}
\end{center}
\caption{$-\omega_{I}$ vs. the mass.}
\label{FreqTable2}
\end{figure}



\section{Neutron Stars}

We have worked with the simplest model available for neutron stars, 
that of a noninteracting Fermi gas of neutrons. For those, the maximum 
mass is $0.72M_{s}$ \cite{Haensel}, and we have selected a few values 
of neutron star masses to work with. For a brief description of the 
EOS involved, see \cite{Weinberg}. Here we shall work on the QNMs 
and other features of the stellar perturbations. The masses we 
have selected for comparison with other kinds of stars were $M=1048m$, 
$M=977m$, $M=665m$ and $M=330m$. Such values are simply a matter of 
choice, having no special feature. They simply correspond to some 
particular choices of central density, $\varepsilon_{0}$, namely 
$4.0\cdot 10^{15}g/cm^3$, $1.5\cdot 10^{15}g/cm^3$, $3.0\cdot 
10^{14}g/cm^3$ and $5.0\cdot 10^{13}g/cm^3$, respectively. These 
values, in turn, are also just a matter of choice. Further values 
could have been equally taken.

\begin{figure}[h]
\begin{center}
\rotatebox{-90}{\mbox{\epsfig{file=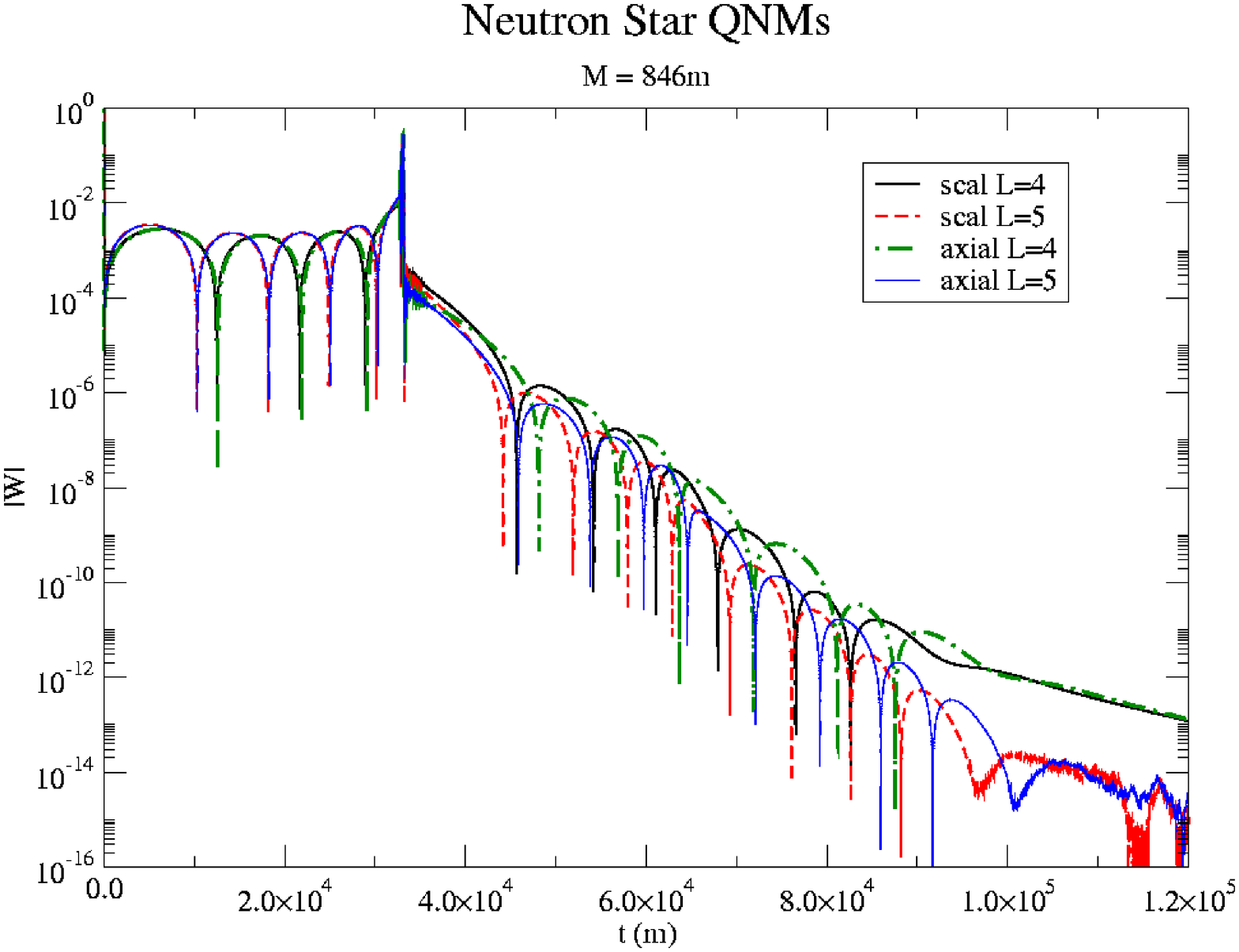,width=0.6\linewidth,clip=}}}
\end{center}
\caption{QNMs for the scalar and axial fields when $M=846m$, $\ell=4,5$. }
\label{MQNnstar1}
\end{figure}

\begin{figure}[h]
\begin{center}
\rotatebox{-90}{\mbox{\epsfig{file=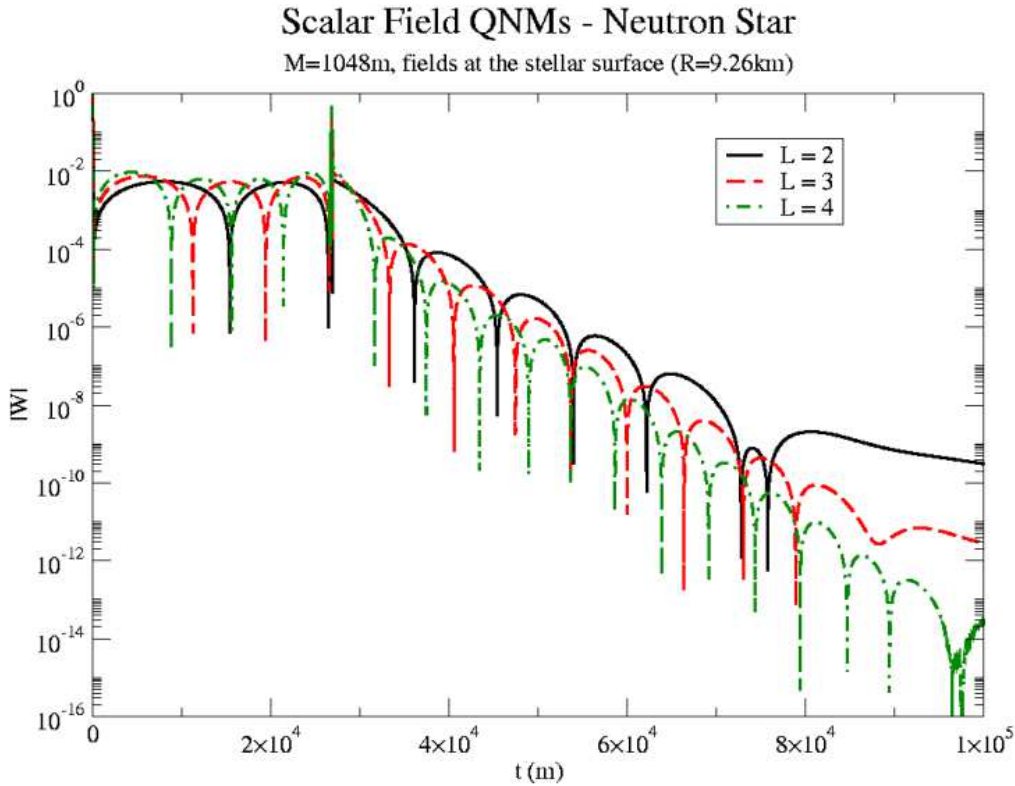,width=0.72\linewidth,clip=}}}
\end{center}
\caption{Scalar field QNMs when $M=1048m$, with changing $\ell$. 
As one would expect, fields oscillate faster for higher $\ell$.}
\label{MQNnstar4}
\end{figure}

\begin{figure}[h]
\begin{center}
\rotatebox{-90}{\mbox{\epsfig{file=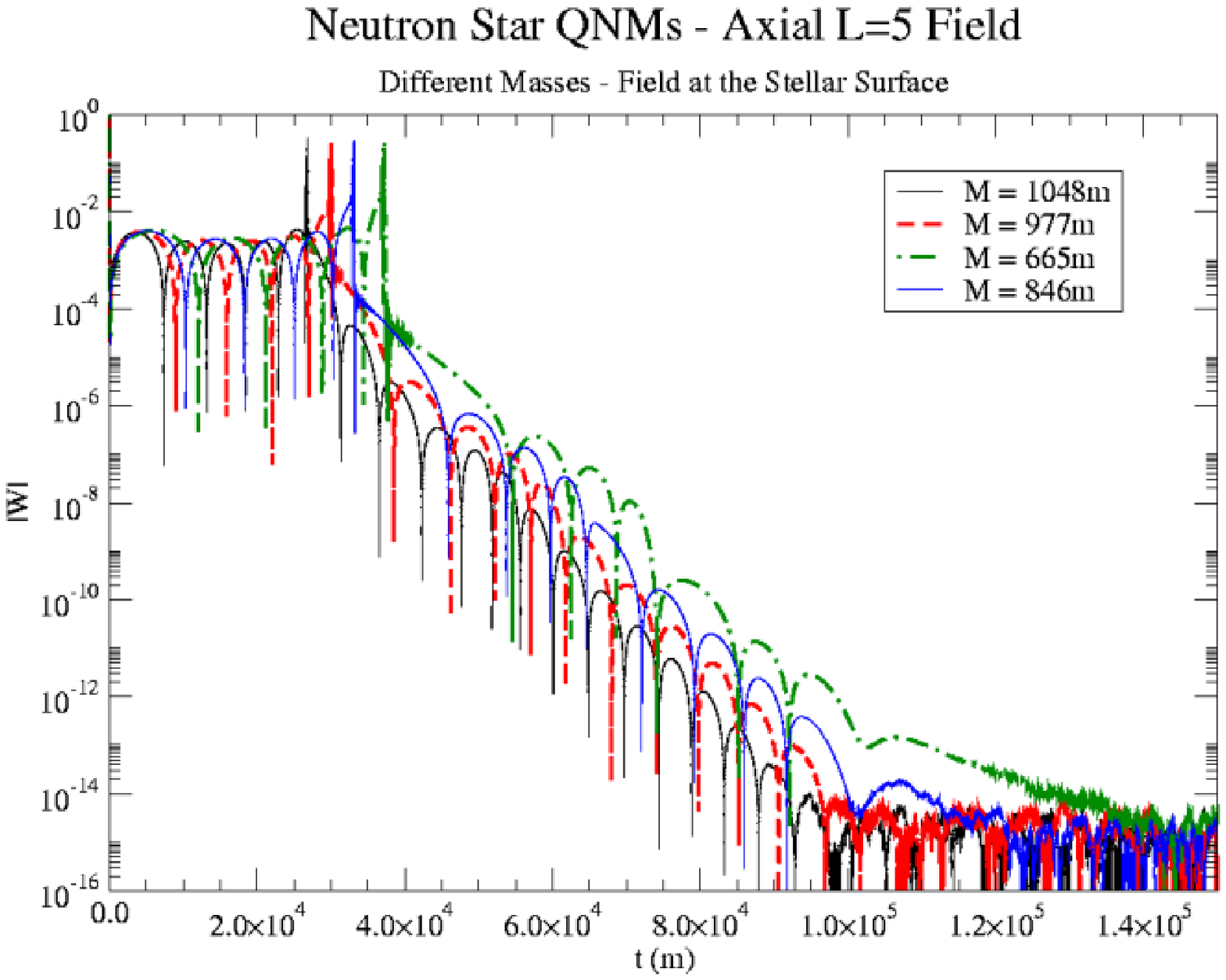,width=0.72\linewidth,clip=}}} 
\end{center}
\caption{$\ell=5$ axial field, seen at the stellar surface. Several
 masses for neutron stars. Notice the increase in the oscillation rate with increasing $M$.}
\label{MQNnstar5}
\end{figure}

Some of our results are available in Figs. (\ref{MQNnstar1}), 
(\ref{MQNnstar4}) and (\ref{MQNnstar5}). 


It would be interesting to compare these results to those coming from
the Schwarzschild black hole QNM analysis. Some of our data on neutron
stars can be seen in tables (\ref{TabModesNstar2}) and
(\ref{TabModesNstar3}). These tables also display $M\omega$, and upon
examining them we verify that the
simple mass-frequency scaling property seen in the Schwarzschild BH and in the
uniform star contexts did not show up here. The product $M\omega$
decreases with decreasing $M$, both for its real part and for the
negative of its imaginary part.


\begin{table}
\begin{tabular}{|c|c|c|c|}
\hline 
$M(m)$ & $\ell$ & $\omega(\times 10^{-4})$ & $M\omega$ \\
\hline
$1048.25$ & $2$ & $3.78-2.86i$ & $0.396-0.300i$ \\
\hline
$1048.25$ & $3$ & $4.86-3.15i$ & $0.509-0.330i$ \\
\hline
$1048.25$ & $4$ & $5.91-3.58i$ & $0.619-0.375i$ \\
\hline
$1048.25$ & $5$ & $7.06-3.94i$ & $0.740-0.413i$ \\
\hline
$977.12$ & $2$ & $-$ & $-$ \\
\hline
$977.12$ & $3$ & $4.04-3.18i$ & $0.395-0.311i$ \\
\hline
$977.12$ & $4$ & $4.78-3.41i$ & $0.467-0.333i$ \\
\hline
$977.12$ & $5$ & $5.64-3.45i$ & $0.551-0.337i$ \\
\hline
$846.54$ & $2$ & $-$ & $-$ \\
\hline
$846.54$ & $3$ & $3.55-2.79i$ & $0.300-0.236i$ \\
\hline
$846.54$ & $4$ & $4.05-3.28i$ & $0.343-0.277i$ \\
\hline
$846.54$ & $5$ & $4.86-3.31i$ & $0.411-0.280i$ \\
\hline 
\end{tabular}
\caption{Frequency data on Neutron Stars, several masses, 
scalar field. Dashes indicate lack of reliable data. The 
$\ell=0,1$ modes showed very few oscillations before decaying to a power-law tail.}
\label{TabModesNstar2}
\end{table}

\begin{table}[tbp]
\begin{tabular}{|c|c|c|c|}
\hline
$Mass(m)$ & $\ell$ & $\omega^{surf}(\times 10^{-4})$ & $M\omega$ \\
\hline
$1048.25$ & $2$ & $2.81-2.49i$ & $0.294-0.261i$ \\
\hline
$1048.25$ & $3$ & $4.23-2.76i$ & $0.443-0.289i$ \\
\hline
$1048.25$ & $4$ & $5.61-3.45i$ & $0.588-0.362i$ \\
\hline
$1048.25$ & $5$ & $6.65-3.44i$ & $0.697-0.361i$ \\
\hline
$977.12$ & $2$ & $-$ & $-$ \\
\hline
$977.12$ & $3$ & $-$ & $-$ \\
\hline
$977.12$ & $4$ & $4.42-3.25i$ & $0.432-0.318i$ \\
\hline
$977.12$ & $5$ & $5.70-3.84i$ & $0.557-0.372i$ \\
\hline
$846.54$ & $2$ & $-$ & $-$ \\
\hline
$846.54$ & $3$ & $-$ & $-$ \\
\hline
$846.54$ & $4$ & $3.65-2.96i$ & $0.309-0.250i$ \\
\hline
$846.54$ & $5$ & $4.42-3.04i$ & $0.374-0.257i$ \\
\hline
\end{tabular}
\caption{Frequency data on neutron stars. Axial field, different masses.}
\label{TabModesNstar3}
\end{table}


The corresponding frequencies for the Schwarzschild BH are available 
in Table (\ref{kompmqn1}). We have picked, for these, masses similar 
to those used for the neutron stars, and the data were taken from 
\cite{dgiugno-06}.

\begin{table}[tbp]
\begin{tabular}{|c|c|c|}
\hline
$field$ & $\ell$ & $M\omega$ \\ 
\hline
$scalar$ & $2$ & $0.483644-0.0967590i$ \\ 
\hline 
$axial$ & $2$ & $0.37367-0.08896i$  \\ 
\hline
$scalar$ & $3$ & $0.675367-0.0964997i$  \\ 
\hline
$axial$ & $3$ & $0.599444-0.0927031i$  \\ 
\hline
$scalar$ & $4$ & $0.867417-0.0963923i$  \\ 
\hline
$axial$ & $4$ & $0.809180-0.0941643i$  \\ 
\hline
$scalar$ & $5$ & $1.059614-0.0963337i$  \\ 
\hline
$axial$ & $5$ & $1.012297-0.0948713i$  \\ 
\hline
\end{tabular}
\caption{{\protect\small \textit{QNM frequencies for the Schwarzschild
BH, for the sake of comparison. We have chosen to show $M\omega$ due
to the mass-frequency scaling property of such Schwarzschild BH modes.}}}
\label{kompmqn1}
\end{table}

The tables (\ref{TabModesNstar2}) and (\ref{TabModesNstar3}) show 
the stellar QNM frequencies
obtained at the stellar surface. We must mention that the
location of extraction - in the time domain - of the modes mattered in
the final result, although it is not visible in tables 
(\ref{TabModesNstar2}) and (\ref{TabModesNstar3}),
because we have taken average values, that is, the (arithmetic) average of several
values measured at different time intervals. For example, for the case
of a neutron star with $M=1048m$ and with an $\ell=4$ axial
peturbation, we have gotten $\omega_{1}=5.55\cdot 10^{-4}-3.45\cdot
10^{-4}i$, $\omega_{2}=5.47\cdot 10^{-4}-3.32\cdot 10^{-4}i$ and
$\omega_{3}=5.53\cdot 10^{-4}-3.39\cdot 10^{-4}i$ for the time
intervals $t_{1}=60000-90000m$, $t_{2}=65000-85000m$ and
$t_{3}=61000-91000m$, respectively, yielding an average value of
$\omega=5.52\cdot 10^{-4}-3.39\cdot 10^{-4}i$. Similar variations 
were also seen for other $M,\ell$ values and fields. A similar phenomenon
also happened for quark stars, although in a less pronounced way. 
No clear overtones were found in the neutron star context, while 
- in a few cases only - we could clearly find them in the quark 
star context (see next section).


From tables (\ref{TabModesNstar2}) and (\ref{TabModesNstar3}), we 
can draw a few conclusions. First, the axial modes have oscillated 
slowlier than their scalar counterparts, exactly as in the Schwarzschild 
BH case. But what has caught our attention, in fact, was the behavior of 
the oscillation rate ($Re(\omega)$) of the modes as a function of the 
stellar mass: the smaller the latter, the smaller the former, contrary 
to what is observed in the Schwarzschild case. More data are needed to 
clarify the matter. 

Upon comparing the tables (\ref{TabModesNstar2}) and (\ref{TabModesNstar3}) 
to table (\ref{kompmqn1}), we may say that neutron star QNMs (at least in 
this simplified model we have dealt with) oscillate more slowly and decay 
considerably faster than Schwarzschild's, at least for the few cases which 
we were able to study.

\section{Quark Stars}

In what follows, we deal with the issue of searching for and analysing
the QNMs of very simple quark stars. The ones we have searched for
obeyed a very simple EOS - the so-called MIT Bag Model - in which the pressure $p$ is given by
\begin{equation}
p=\frac{1}{3}(\varepsilon-4B),
\end{equation}
where $B$ stands for the bag constant. For more details on this model
see \cite{Weber04}, \cite{Glendenning} and references therein. For
further models of quark stars, see \cite{Glendenning} and \cite{Su}. 

Upon dealing with the present class of stars, first of all, we have searched for 
values of $\varepsilon_{0}$ and $B$ which yielded masses very near those we had 
gotten for our neutron stars. For those, our limiting mass was around $0.72M_{s}$ 
and, according to \cite{Witten}, there is an empirical expression for the maximum 
mass for a given $B$, namely,
\begin{equation}
M_{max}=\frac{1.96M_{s}}{\sqrt{B/B_{c}}},
\end{equation}
where $B_c = 60 MeV/fm^{3}$. In order to have $M_{max}= 0.72 M_s$, one must choose
 $B=445MeV/fm^3$. Once we have found that, we just determine - by trial and 
error - the central densities $\varepsilon_{0}$ which yield the masses we want. 

Once the stars were integrated, we had all the data on the perturbative potentials, 
as we did for the uniform and neutron stars, and the QNM-searching routine could be 
started to find the modes. 

\begin{figure}[h]
\begin{center}
\rotatebox{-90}{\mbox{\epsfig{file=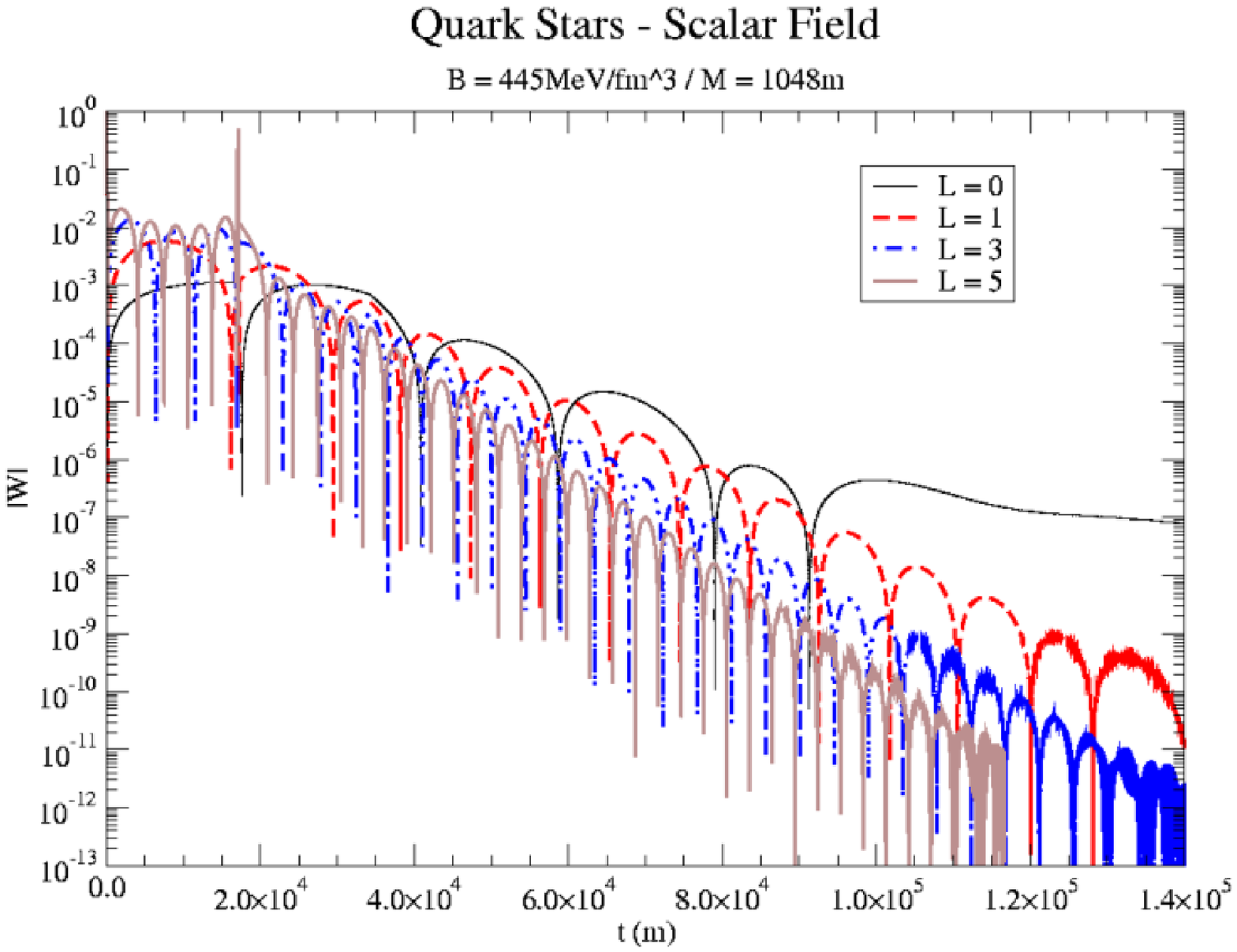,width=0.72\linewidth,clip=}}}
\end{center}
\caption{Scalar Field QNMs, Quark Stars, $B=445MeV/fm^3$ and $M=1048m$.}
\label{QStarGraph1}
\end{figure}

\begin{figure}[h]
\begin{center}
\rotatebox{-90}{\mbox{\epsfig{file=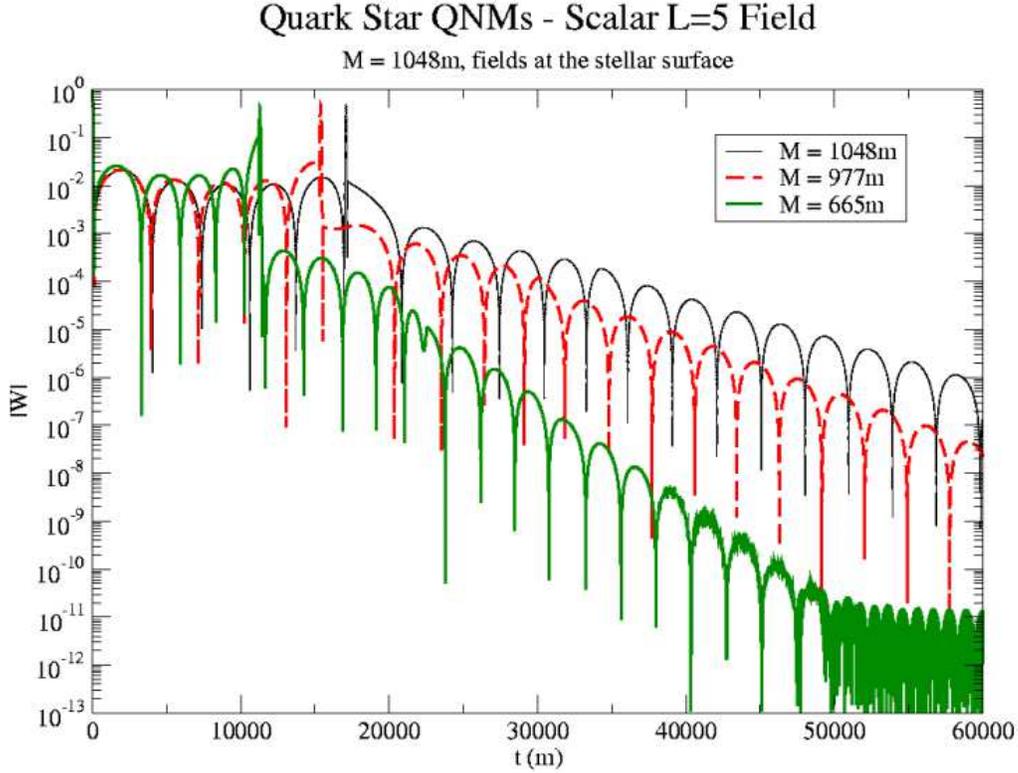,width=0.72\linewidth,clip=}}} 
\end{center}
\caption{Scalar $\ell=5$ Field QNMs, Quark Stars, $B=445MeV/fm^3$ and different masses.}
\label{QStarGraph2}
\end{figure}

\begin{figure}[h]
\begin{center}
\rotatebox{-90}{\mbox{\epsfig{file=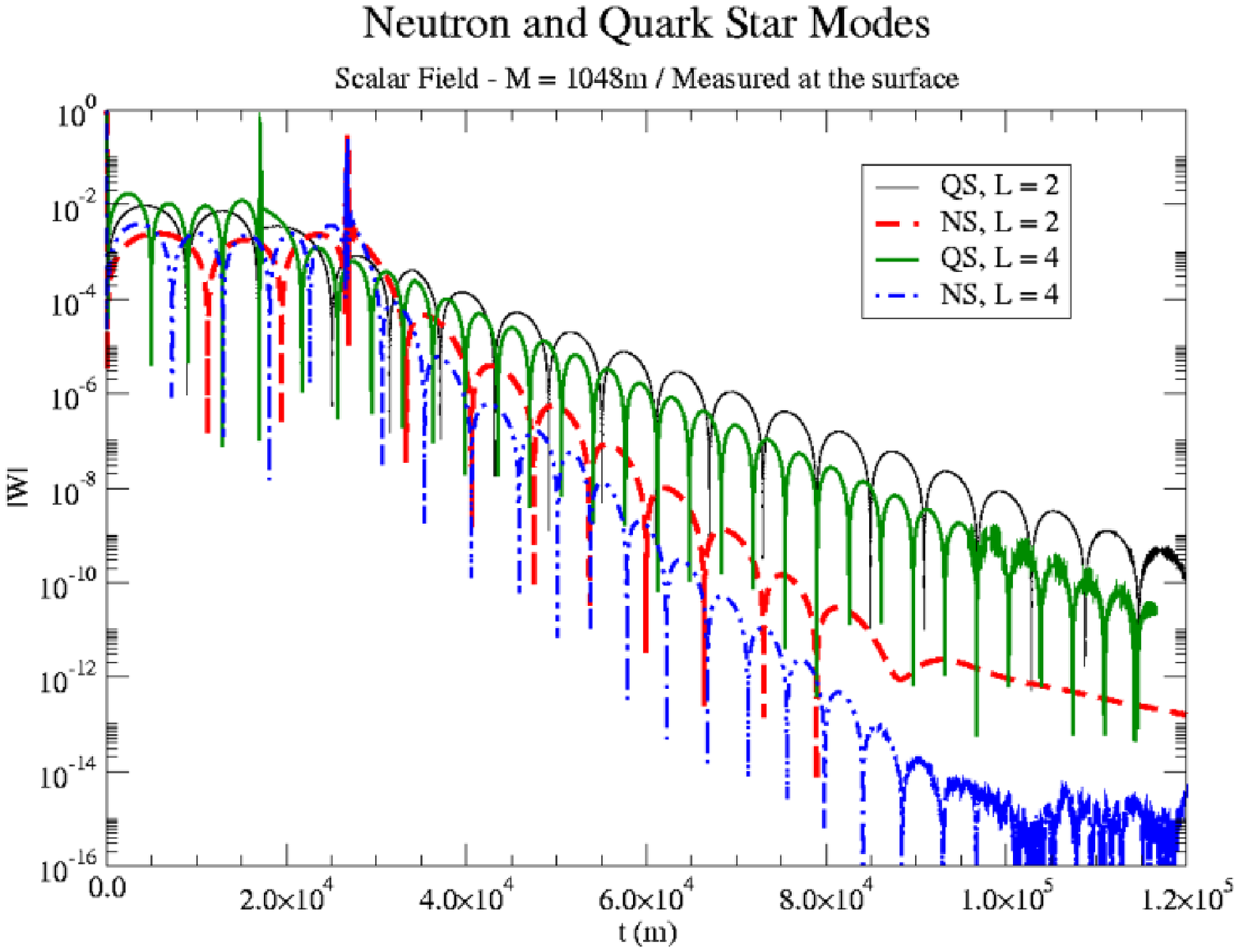,width=0.72\linewidth,clip=}}}\end{center}
\caption{Comparing QNMs from Neutron and Quark Stars, $M=1048m$.}
\label{CompNSQS1}
\end{figure}

For these stars, we have some results, which can be seen in Figs. (\ref{QStarGraph1}) 
and (\ref{QStarGraph2}). A comparison between quark and neutron star modes is provided 
in Fig. (\ref{CompNSQS1}).

The figures above show that quark star modes, for a given mass, field type and $\ell$, 
oscillate considerably faster than neutron star modes, besides being less damped. 
Recall that the most massive neutron stars in this very simple model have a maximum 
mass of $1048m\thickapprox 0.72M_{s}$, and that they experience an increase in their 
oscillation rates for higher masses - just the opposite of quark stars. Thus, for 
any mass below this limit, neutron star modes will oscillate less than quark star modes.

\begin{table}[tbp]
\begin{tabular}{|c|c|c|c|c|}
\hline
$M(m)$ & $\ell$ & $\omega^{surf}(\times 10^{-4})$ & $M\omega$ \\
\hline
$1048$ & $1$ & $3.47-1.45i$ & $0.364-0.152i$ \\
\hline
$1048$ & $2$ & $5.26-1.63i$ & $0.551-0.171i$ \\
\hline
$1048$ & $3$ & $7.06-1.78i$ & $0.740-0.187i$ \\
\hline
$1048$ & $4$ & $8.85-1.92i$ & $0.927-0.201i$ \\
\hline
$1048$ & $5$ & $10.6-2.06i$ & $1.11-0.216i$ \\
\hline
$977$ & $1$ & $3.58-1.84i$ & $0.350-0.180i$ \\
\hline
$977$ & $2$ & $5.42-2.09i$ & $0.530-0.204i$ \\
\hline
$977$ & $3$ & $7.26-2.27i$ & $0.709-0.220i$ \\
\hline
$977$ & $4$ & $9.11-2.45i$ & $0.890-0.239i$ \\
\hline
$977$ & $5$ & $10.9-2.62i$ & $1.06-0.254i$ \\
\hline
$846$ & $1$ & $3.76-2.38i$ & $0.318-0.201i$ \\
\hline
$846$ & $2$ & $5.77-2.90i$ & $0.488-0.245i$ \\
\hline
$846$ & $3$ & $7.75-3.04i$ & $0.656-0.257i$ \\
\hline
$846$ & $4$ & $9.79-3.13i$ & $0.828-0.265i$ \\
\hline
$846$ & $5$ & $11.7-3.58i$ & $0.990-0.303i$ \\
\hline
\end{tabular}
\caption{Frequency data for $B=445MeV/fm^3$ Quark Stars, several masses, 
scalar field. Dashes indicate the lack of reliable data.}
\label{QS20}
\end{table}

As for our current data on the axial modes, they seemed not to be of good 
quality and we have yet to find out why.

\begin{figure}[h]
\begin{center}
\rotatebox{-90}{\mbox{\epsfig{file=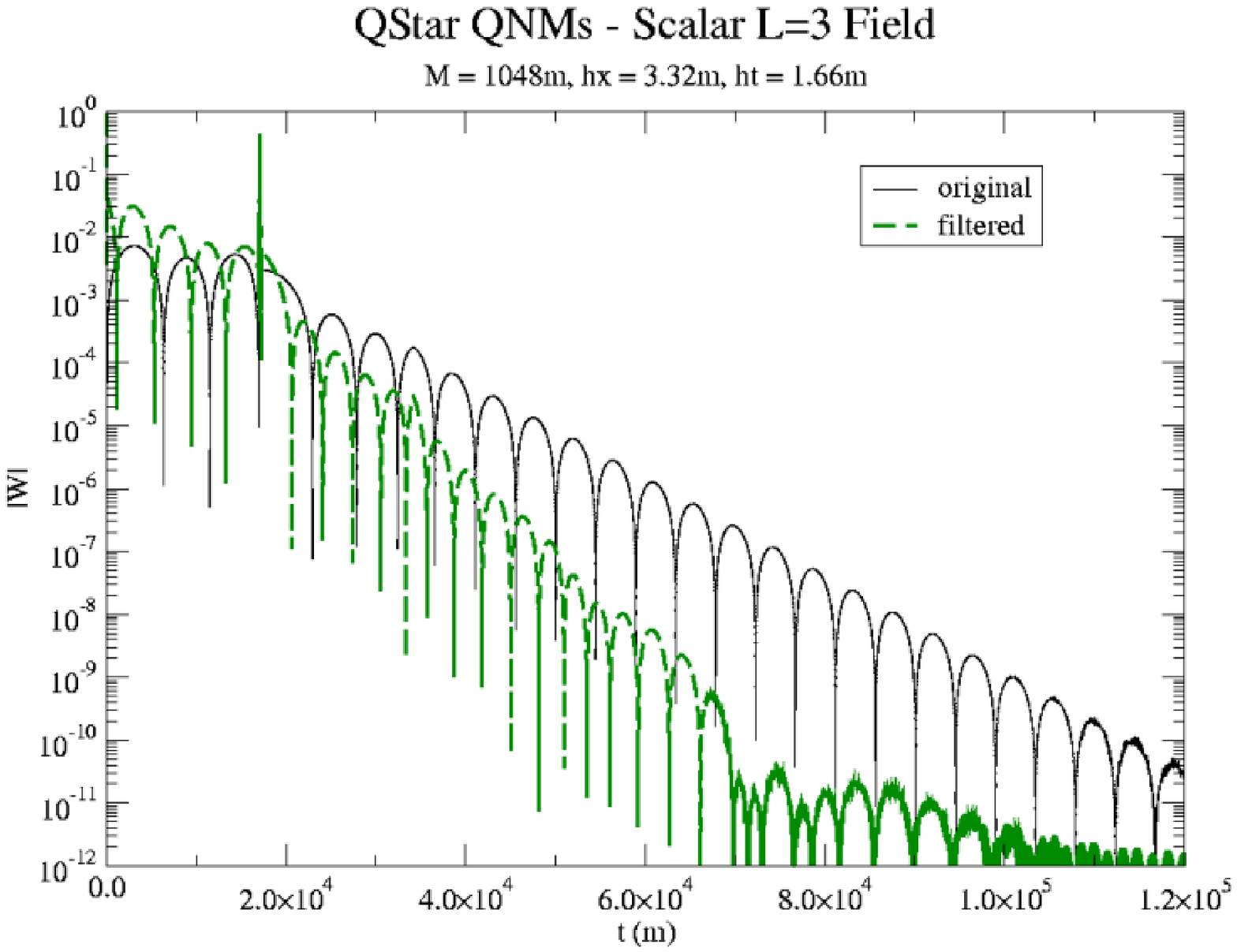,width=0.72\linewidth,clip=}}} 
\end{center}
\caption{Scalar $\ell=3$ field, $M=1048m$. Notice the appearance of the 
first overtone, even if it looks somewhat irregular.}
\label{QStarGraph3}
\end{figure}

One point should be stressed about the data in Table (\ref{QS20}): as
we had already commented on the neutron star context, depending on the
region of the time domain used to make the fittings to find the QNM
frequencies, the results differed a bit (not as much as for the
neutron stars). We have decided to rely on the data obtained at the
end of the time domain - because any overtone tends to show itself at
the beginning of the wave profile - but before the onset of any
instability (seen by the 'trembling' data). After making such a
fitting, we decided to subtract it from our original function (as we
had done in \cite{dgiugno-06}, in another context). To our surprise,
we could find a secondary mode in most cases, though only in a few
cases it was possible to make a minimally reliable fitting to it. See
Fig. (\ref{QStarGraph3}). 

An extra remark is due on the data in Table (\ref{QS20}): we computed
also $M\omega$, and we have not seen the same simple scaling property
we had seen before for the Schwarzschild BH modes and for the uniform
stars. The real part of $M\omega$ decreases with $M$, and the opposite
holds for the imaginary part of $-M\omega$.

\begin{figure}[h]
\begin{center}
\rotatebox{-90}{\mbox{\epsfig{file=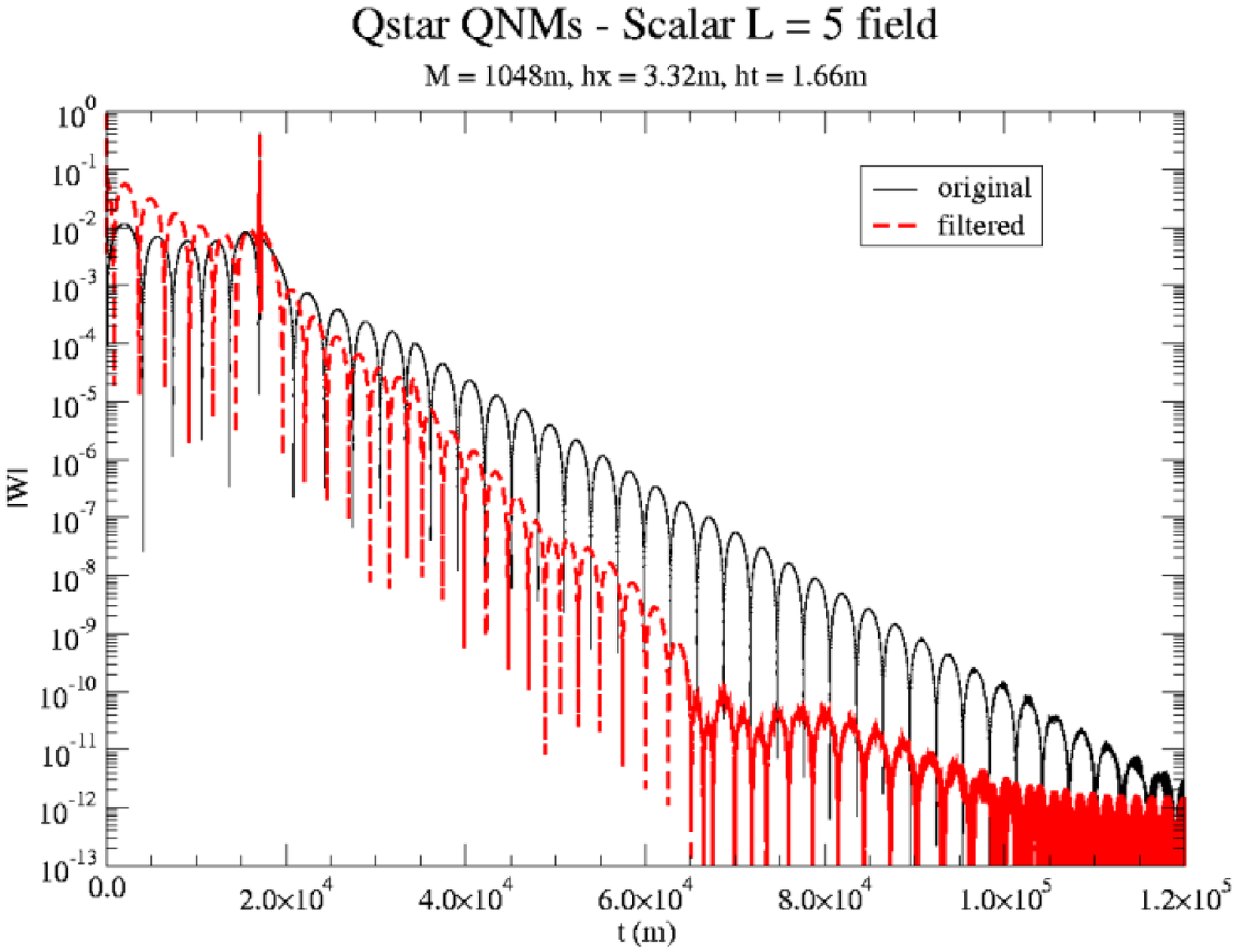,width=0.72\linewidth,clip=}}} 
\end{center}
\caption{Scalar $\ell=5$ field, $M=1048m$. Notice the first overtone.}
\label{QStarGraph4}
\end{figure}

In Fig. (\ref{QStarGraph3}), for example, one has a secondary mode with 
$\omega=0.00105-0.000400i$, compared to $\omega=0.000706-0.000178i$ for 
the fundamental mode. For the scalar $\ell=5$ field, in Fig. (\ref{QStarGraph4}), 
the figures are $\omega=0.00135-0.000364i$ and $0.00106-0.000206i$, respectively. 
That is, overtones were a recurrent theme in our quest for stellar modes.

The oscillations become slower as the masses increase, and so does 
the damping rates, much as in the Schwarzschild BH case, and contrary 
to the neutron star case.
\section{Comparative Charts}

In this section we compare all kinds of stars mentioned in this paper
between themselves and to Schwarzschild BHs. 

\begin{table}
\begin{tabular}{|c|c|c|c|c|}
\hline
$Mass$ & $Star/BH$ & $\ell$ & $\omega_{R}$ & $-\omega_{I}$ \\
\hline
$1048.25$ & $NS$ & $2$ & $3.78E-4$ & $2.86E-4$ \\
\hline
$1048.25$ & $NS$ & $3$ & $4.86E-4$ & $3.15E-4$ \\
\hline
$1048.25$ & $NS$ & $4$ & $5.91E-4$ & $3.58E-4$ \\
\hline
$1048.25$ & $NS$ & $5$ & $7.06E-4$ & $3.94E-4$ \\
\hline
$1048.25$ & $QS$ & $1$ & $3.47E-4$ & $1.45E-4$ \\
\hline
$1048.25$ & $QS$ & $2$ & $5.26E-4$ & $1.63E-4$ \\
\hline
$1048.29$ & $QS$ & $3$ & $7.06E-4$ & $1.78E-4$ \\
\hline
$1048.29$ & $QS$ & $4$ & $8.85E-4$ & $1.92E-4$ \\
\hline
$1048.25$ & $QS$ & $5$ & $1.06E-3$ & $2.06E-4$ \\
\hline
$1048.25$ & $SCH$ & $3$ & $6.4428E-4$ & $9.206E-5$ \\
\hline
$1048.25$ & $SCH$ & $4$ & $8.2749E-4$ & $9.196E-5$ \\ 
\hline
$1048.25$ & $SCH$ & $5$ & $1.0108E-3$ & $9.190E-5$ \\
\hline
$977.12$ & $NS$ & $1$ & $-$ & $-$ \\
\hline
$977.12$ & $NS$ & $2$ & $-$ & $-$ \\
\hline
$977.12$ & $NS$ & $3$ & $4.04E-4$ & $3.18E-4$ \\
\hline
$977.12$ & $NS$ & $4$ & $4.78E-4$ & $3.41E-4$ \\
\hline 
$977.12$ & $NS$ & $5$ & $5.64E-4$ & $3.45E-4$ \\
\hline
$977.12$ & $QS$ & $1$ & $3.58E-4$ & $1.84E-4$ \\
\hline
$977.12$ & $QS$ & $2$ & $5.42E-4$ & $2.09E-4$ \\
\hline
$977.12$ & $QS$ & $3$ & $7.26E-4$ & $2.27E-4$ \\
\hline
$977.12$ & $QS$ & $4$ & $9.11E-4$ & $2.45E-4$ \\
\hline 
$977.12$ & $QS$ & $5$ & $1.09E-3$ & $2.62E-4$ \\
\hline
$977.12$ & $SCH$ & $1$ & $2.9980E-4$ & $9.9953E-5$ \\
\hline
$977.12$ & $SCH$ & $2$ & $4.9497E-4$ & $9.9025E-5$ \\
\hline
$977.12$ & $SCH$ & $3$ & $6.9118E-4$ & $9.8759E-5$ \\
\hline
$977.12$ & $SCH$ & $4$ & $8.8773E-4$ & $9.8649E-5$ \\
\hline
$977.12$ & $SCH$ & $5$ & $1.0844E-3$ & $9.8589E-5$ \\
\hline
\end{tabular}
\caption{Comparing neutron stars (NS) to quark stars (QS) to
  Schwarzschild black holes (SCH), scalar field. Fields measured at
  the stellar surface, for stars.}
\label{CC1}
\end{table}

\begin{table}
\begin{tabular}{|c|c|c|c|c|}
\hline
$846.54$ & $NS$ & $1$ & $-$ & $-$ \\
\hline
$846.54$ & $NS$ & $2$ & $-$ & $-$ \\
\hline
$846.54$ & $NS$ & $3$ & $3.55E-4$ & $2.79E-4$ \\
\hline
$846.54$ & $NS$ & $4$ & $4.05E-4$ & $3.28E-4$ \\
\hline 
$846.54$ & $NS$ & $5$ & $4.86E-4$ & $3.31E-4$ \\
\hline
$846.54$ & $QS$ & $1$ & $3.76E-4$ & $2.38E-4$ \\
\hline
$846.54$ & $QS$ & $2$ & $5.77E-4$ & $2.90E-4$ \\
\hline
$846.54$ & $QS$ & $3$ & $7.75E-4$ & $3.04E-4$ \\
\hline
$846.54$ & $QS$ & $4$ & $9.79E-4$ & $3.13E-4$ \\
\hline 
$846.54$ & $QS$ & $5$ & $1.17E-3$ & $3.58E-4$ \\
\hline
$846.54$ & $SCH$ & $1$ & $3.4604E-4$ & $1.1537E-4$ \\
\hline
$846.54$ & $SCH$ & $2$ & $5.7132E-4$ & $1.1430E-4$ \\
\hline
$846.54$ & $SCH$ & $3$ & $7.9780E-4$ & $1.1399E-4$ \\
\hline
$846.54$ & $SCH$ & $4$ & $1.0247E-3$ & $1.1387E-4$ \\
\hline
$846.54$ & $SCH$ & $5$ & $1.2517E-3$ & $1.1380E-4$ \\
\hline
\end{tabular}
\caption{Continuation of the previous table.}
\label{CC1b}
\end{table}

\begin{table}
\begin{tabular}{|c|c|c|c|c|}
\hline
$Mass$ & $Star/BH$ & $\ell$ & $\Omega_{R}$ & $-\Omega_{I}$ \\
\hline
$1048.25$ & $NS$ & $2$ & $2.81E-4$ & $2.49E-4$ \\
\hline
$1048.25$ & $NS$ & $3$ & $4.23E-4$ & $2.49E-4$ \\
\hline
$1048.25$ & $NS$ & $4$ & $5.61E-4$ & $3.45E-4$ \\
\hline
$1048.25$ & $NS$ & $5$ & $6.65E-4$ & $3.44E-4$ \\
\hline
$1048.25$ & $SCH$ & $2$ & $3.5647E-4$ & $8.4865E-5$ \\
\hline
$1048.25$ & $SCH$ & $3$ & $5.7185E-4$ & $8.8436E-5$ \\
\hline
$1048.25$ & $SCH$ & $4$ & $7.7193E-4$ & $8.9829E-5$ \\
\hline
$1048.25$ & $SCH$ & $5$ & $9.6570E-4$ & $9.0504E-5$ \\
\hline
$977.12$ & $NS$ & $2$ & $-$ & $-$ \\
\hline
$977.12$ & $NS$ & $3$ & $3.25E-4$ & $2.46E-4$ \\
\hline
$977.12$ & $NS$ & $4$ & $4.42E-4$ & $3.25E-4$ \\
\hline
$977.12$ & $NS$ & $5$ & $5.70E-4$ & $3.84E-4$ \\
\hline
$977.12$ & $SCH$ & $2$ & $3.8242E-4$ & $9.1043E-5$ \\
\hline
$977.12$ & $SCH$ & $3$ & $6.1348E-4$ & $9.4874E-5$ \\
\hline
$977.12$ & $SCH$ & $4$ & $8.2813E-4$ & $9.6368E-5$ \\
\hline
$977.12$ & $SCH$ & $5$ & $1.0360E-3$ & $9.7093E-4$ \\
\hline
$846.54$ & $NS$ & $2$ & $-$ & $-$ \\
\hline
$846.54$ & $NS$ & $3$ & $-$ & $-$ \\
\hline
$846.54$ & $NS$ & $4$ & $3.64E-4$ & $2.96E-4$ \\
\hline
$846.54$ & $NS$ & $5$ & $4.42E-4$ & $3.04E-4$ \\
\hline
$846.54$ & $SCH$ & $2$ & $4.4141E-4$ & $1.0509E-4$ \\
\hline
$846.54$ & $SCH$ & $3$ & $7.0811E-4$ & $1.0951E-4$ \\
\hline
$846.54$ & $SCH$ & $4$ & $9.5587E-4$ & $1.1123E-4$ \\
\hline
$846.54$ & $SCH$ & $5$ & $1.1958E-3$ & $1.1207E-4$ \\
\hline
\end{tabular}
\caption{Comparing neutron stars (NS) to
  Schwarzschild black holes (SCH), axial field. Fields measured at
  the stellar surface, for stars.}
\label{CC2}
\end{table}

We did not place $M\omega$ in tables (\ref{CC1}), (\ref{CC1b}) and (\ref{CC2})  
because the neutron and
quark stars do not share the simple $\omega\propto\frac{1}{M}$ property
with the Schwarzschild BHs, as seen in their respective sections.   
From the aforementioned tables we learn that the neutron stars modes are more
strongly damped than its quark counterparts and these are, in turn,
more damped than Schwarzschild modes. The Schwarzschild modes
oscillate faster than the neutron star modes, but slightly less than the quark star modes.

\begin{figure}[h]
\begin{center}
\rotatebox{-90}{\mbox{\epsfig{file=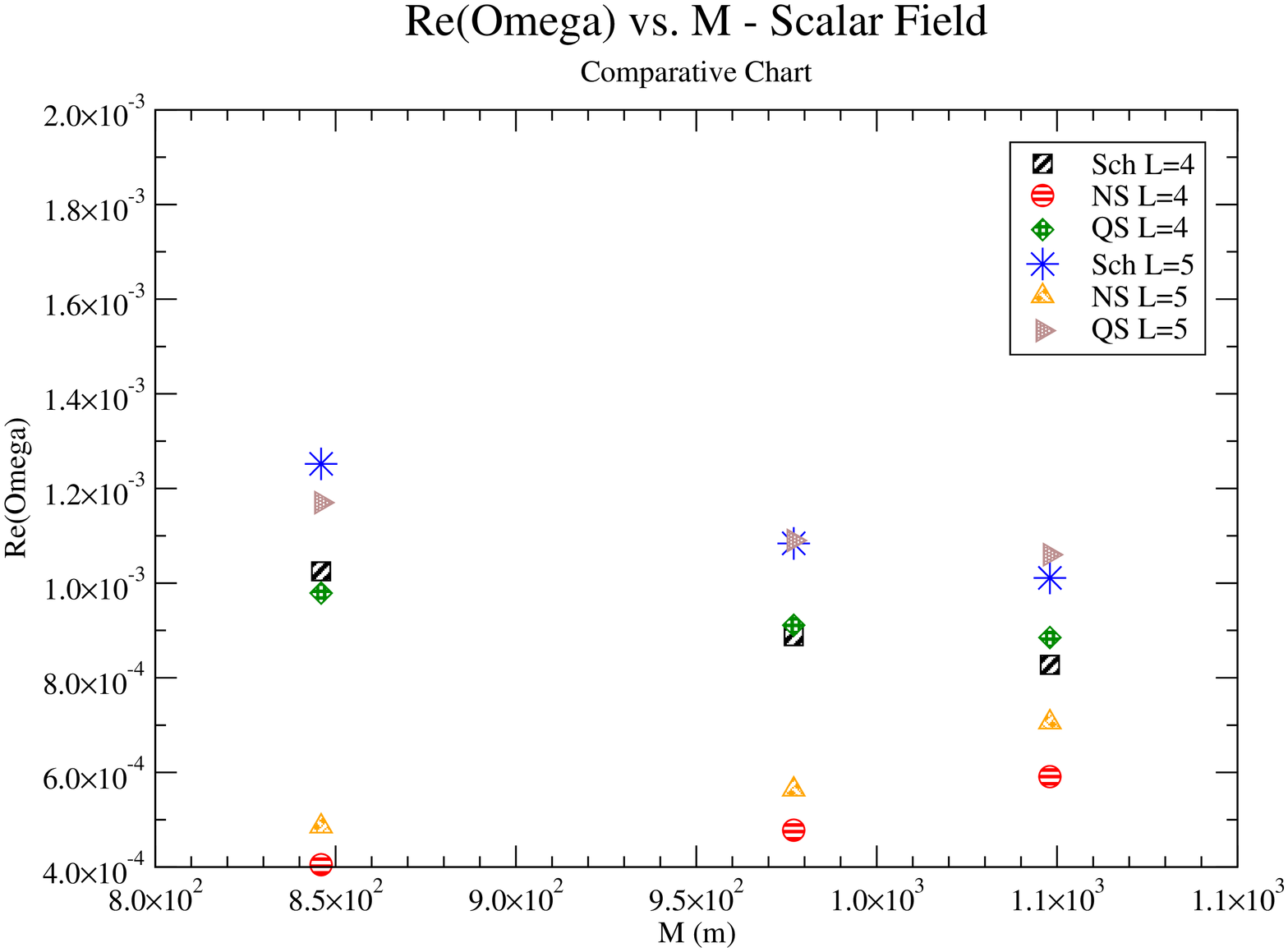,width=0.55\linewidth,clip=}}} 
\end{center}
\caption{Comparing Frequencies. Real Part. Scalar Field.}
\label{X1}
\end{figure}

\begin{figure}[h]
\begin{center}
\rotatebox{-90}{\mbox{\epsfig{file=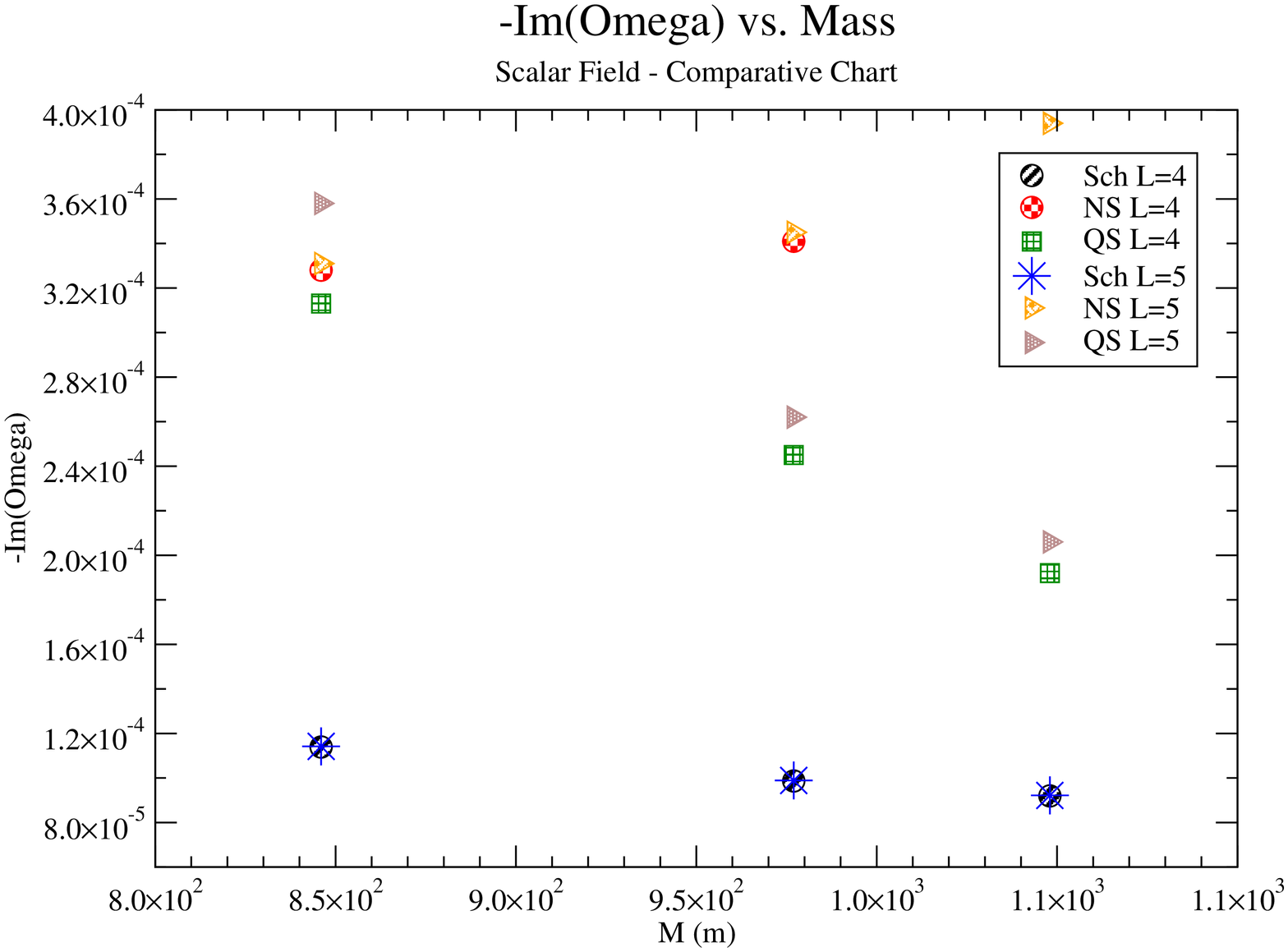,width=0.55\linewidth,clip=}}} 
\end{center}
\caption{Comparing Frequencies. Imaginary Part. Scalar Field.}
\label{X2}
\end{figure}

\section{Final Remarks and Conclusions}

First of all, the oscillating frequencies of the QNMs ($\omega_{R}$)
increases with increasing $\ell$, for a given mass and perturbation, as 
in the black-hole case, irrespective of the star type. The damping 
$-\omega_{I}$ of the modes, however, decreases with increasing $\ell$, 
and does so very markedly, for all the masses and fields under study 
for uniform stars,
whereas in the Schwarzschild black-hole (SchBH) case $-\omega_{I}$ has 
a very slight increase instead for the axial field, while also decreasing 
(very slightly, though) for the scalar field. This is an interesting 
contrast. And uniform stars showed stronger dampings for axial fields 
than for scalar fields, also contrary to SchBH.

The neutron stars (NS) and quark stars (QS) modes, in general, showed
an increase in $-\omega_{I}$ with increased $\ell$ for scalar
perturbations. For the axial modes, a similar trend could be detected
for NS, and the only possible anomaly could be the $M=1048m$ case, when the
$\ell=4$ mode had practically the same damping ratio as the $\ell=5$
mode, as can be seen at the end of the table (\ref{CC2}).

There is also a clear dependence of the modes on the uniform star mass
mass, for a given compactness and perturbation, as in the
Schwarzschild case: the more massive a uniform star is, the slower is the
oscillation of the field and the weaker is its damping. The same holds for 
the overtones we have found for them. As for the compactness $c$ itself, 
given a fixed mass, an increase in $c$ made both $Re(\omega)$ and $-Im(\omega)$ 
decrease, that is, more compact stars had slower and (much) less damped oscillations.

The case of competing modes (overtones) in stars needed our foremost 
attention, since it seems not to have been mentioned anywhere so far. 
We could detect such competing modes in very compact uniform stars, 
especially when $c\thickapprox 0.77$ onwards. And here comes another 
contrast: while in the Schwarzschild black hole context the overtones 
have oscillated more slowly than the fundamental mode (see \cite{Konoplya03}, 
\cite{dgiugno-06}), for the uniform stars the opposite was true. A similar 
feature was detected for some QS modes in which the presence of overtones 
was detectable, although we could not find clear overtones for NS.

In general, leaving aside the uniform stars (UniS) due to their
dependency on $c$, given some perturbation type, $M$ and $\ell$, QS
modes tend to oscillate a bit faster than SchBH modes for higher
masses and slightly slower for smaller masses, and these, in turn, 
oscillate faster than NS modes. When it comes to damping, though, 
the Sch BH modes are the least damped of all, followed by the QS modes 
(at least twice as damped) and by NS modes (the most damped of all). 
That is, NS modes are the slowest in oscillation rate and the most 
damped. At least, in the mass range under scrutiny.

Nevertheless, if UniS are considered, their modes have slightly higher 
$\omega_{R}$ than SchBHs, but their $-\omega_{I}$ can be much smaller 
than their Sch BH counterparts, so that these uniform stellar modes are, 
in fact, the LEAST damped of all modes, particularly when $c$ increases 
towards its limiting value of $8/9$.  

The Sch BH and the uniform stars have a simple scaling property for 
$\omega$, namely $M\omega=k$, where $k$ is a constant depending on the 
perturbation, on $\ell$ and, for the uniform stars, $c$. Not so for the 
NS and QS modes, where we have not found so simple a correlation between 
$M$ and $\omega$. For QS at least, a decreasing $M$ meant an increasing 
$\omega$ (especially for $\omega_{R}$). For NS, a curious feature emerged: 
the larger $M$ is, the higher $\omega$ is, especially $\omega_{R}$, in sharp 
contrast to what has been seen for SchBH, UniS and QS. At any rate, however, 
in the mass range we have worked with, even the most massive NS have 
presented modes which were slower and more damped than the most massive 
SchBH, UniS and QS modes. We are still in the search for the reason(s) 
for such a behaviour. 

\appendix
\section{Brief Theory of Uniform Stars}

Uniform stars are stars possessing a uniform density
$\varepsilon_{0}$, and references abound in the literature, as in 
\cite{Weinberg}. For these stars, computations are very easy, as we
shall see.

The star mass function is simply
\begin{equation}
m(r)=4\pi\frac{\varepsilon_{0}r^3}{3},
\end{equation}
if $r\leq R$ and $m(r)=M$ if $r>R$. Hence, in terms of the 
radius $R$ and the mass $M$ of the star, one has $\varepsilon_{0}=\frac{3M}{4\pi R^3}$.

The pressure is determined via the Oppenheimer-Volkov equation,
\begin{equation}
\frac{dp}{dr}=-\frac{(p+\varepsilon)(m+4\pi r^3p)}{r^2(1-\frac{2m}{r})},
\end{equation}
and is given by
\begin{equation}
p(r)=\frac{3M}{4\pi R^3}\lbrack\frac{\sqrt{1-\frac{2M}{R}}-
\sqrt{1-\frac{2Mr^2}{R^3}}}{\sqrt{1-\frac{2Mr^2}{R^3}}-3\sqrt{1-\frac{2M}{R}}}\rbrack,
\end{equation}
which approaches zero smoothly as $r\to R$. The $g_{tt}$ term of the metric is given by
\begin{equation}
g_{tt}=-\frac{1}{4}(3\sqrt{1-\frac{2M}{R}}-\sqrt{1-\frac{2Mr^2}{R^3}})^2,
\end{equation}
which reduces to the Schwarzschild $g_{tt}$ for $r>R$. One point is of
utmost importance in what follows: the expression for $p$ becomes
singular when $M/R>4/9$. Since the star pressure cannot be infinite
anywhere, one concludes that there is an upper limit to the degree of
compactness ($c=2M/R$) for any star. In fact, since $M/R$ cannot
exceed $4/9$, no star can have $c>8/9$, so there's a gap in $c$ between 
Schwarzschild black holes and spherically symmetric stars of any kind. 
This limitation stems from General Relativity itself, not from any 
particular stellar model. 

From the existing literature \cite{Chand-Ferr},\cite{Chand-FerrB} 
we have the wave equation satisfied by the scalar and gravitational 
perturbations, namely
\begin{equation}
\frac{\partial^2W}{\partial x^2}-\frac{\partial^2W}{\partial t^2}=VW,
\end{equation}
where $W$ stands for the perturbation amplitude and $V$, the 
perturbative potential. The latter is written as
\begin{eqnarray}
V&=&\frac{1}{4}(3\sqrt{1-\frac{2M}{R}}-\sqrt{1-
\frac{2Mr^2}{R^3}})^2[\frac{\ell(\ell+1)}{r^2}+\frac{2\sigma m(r)}
{r^3}+4\pi(\varepsilon_{0}-p(r))]\nonumber\\
V&=&(1-\frac{2M}{r})[\frac{\ell(\ell+1)}{r^2}+\frac{2\sigma M}{r^3}],
\end{eqnarray}
where $\sigma=1,0,-3$ depending on whether we are dealing with scalar, 
electromagnetic or axial perturbations. The first of these equations 
holds inside the star, and the latter holds outside.

\begin{figure}[h]
\begin{center}
\rotatebox{-90}{\mbox{\epsfig{file=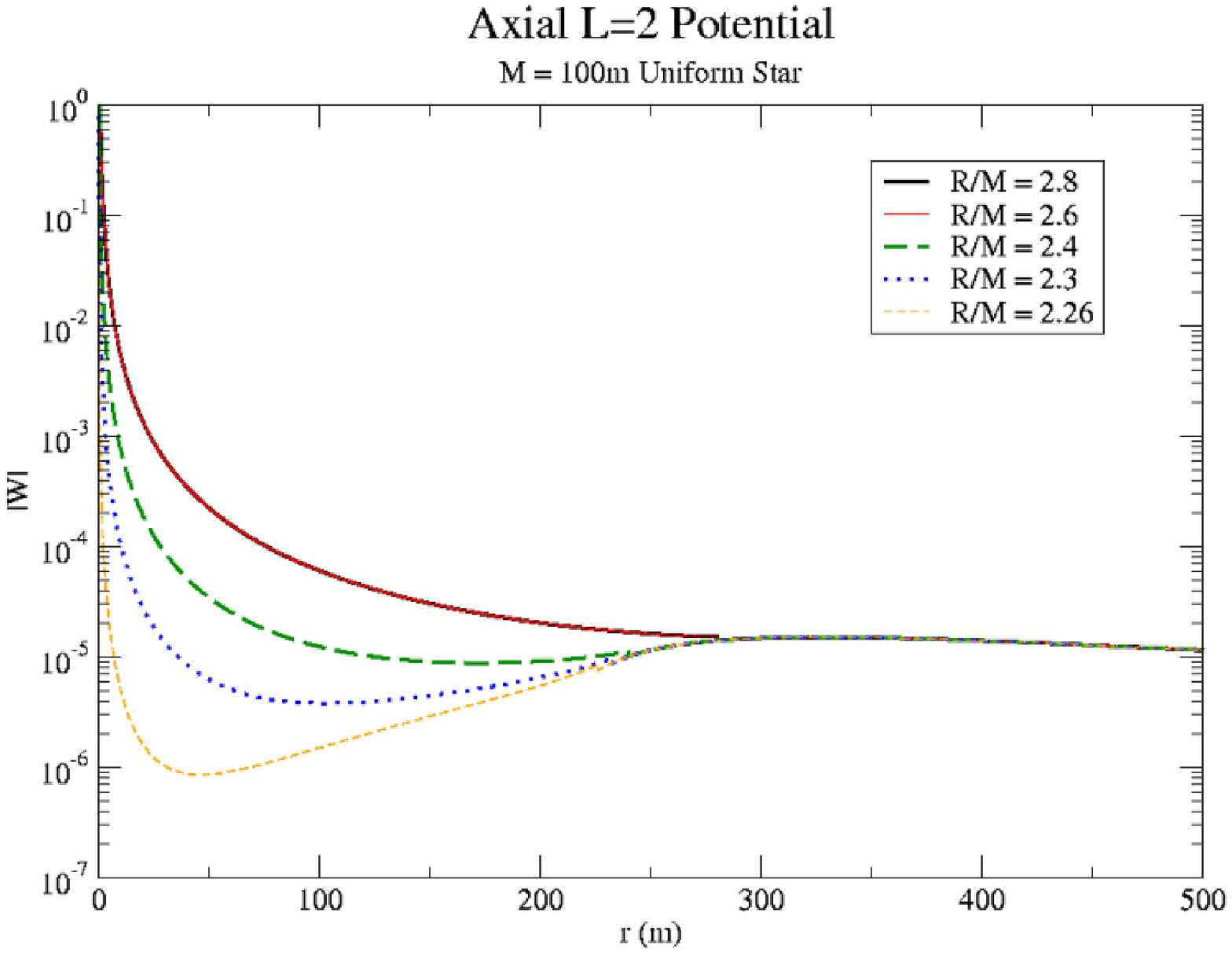,width=0.7\linewidth,clip=}}}
\end{center}
\caption{Graph of de $V$ vs. $r$ for a compact star. Axial $\ell=2$ case shown.}\label{Vstar}
\end{figure}

We can now illustrate some of the potentials, and find out how they
change with the compactness. Such an illustration is available in
Fig. (\ref{Vstar}).

\section{The Numerical Method}

We have employed a direct numerical method consisting of a grid in the
tortise coordinate $x$ and the time coordinate $t$. Since $x$ runs
from some finite $x_{0}$ at $r=0$ to $x\to\infty$ when $r\to\infty$,
we start by specifying the field $W$ (scalar or axial) and its time
derivative at $t=0$ in the region of interest in $x$ (usually a
Gaussian wave packet centered around some $x_{1}>x_{0}$). The time
evolution of the field is given by
\begin{eqnarray}
\Psi(t_{0}+\delta t,x_{0})&=&-\Psi(t_{0}-\delta t,x_{0})+(2-
\delta^2 xV(x_{0})-\frac{5\delta t^2}{2\delta x^2})
\Psi(t_{0},x_{0})+\nonumber \\
&+&\frac{4\delta t^2\lbrack\Psi(t_{0},x_{0}+\delta x)+
\Psi(t_{0},x_{0}-\delta x)\rbrack}{3\delta x^2}-\nonumber \\
&-&\frac{\delta t^2\lbrack\Psi(t_{0},x_{0}-2\delta
  x)+\Psi(t_{0},x_{0}+2\delta x)\rbrack}{12\delta x^2},
\label{xtgridf}
\end{eqnarray}
in which $\delta x$ is the spacing in $x$ and $\delta t$ is the time
step. Given the ratio $m=\frac{\delta t}{\delta x}$ (the so-called
mesh ratio), one must have $m<1$ for the sake of convergence. We stop
at some $t>t_{0}$ and analyse the data for all $x$ at that
time. Reflection may occur at the borders of the grid.


\end{document}